\documentclass[11pt,a4paper]{article}
\pdfoutput=1
\usepackage{jcappub}

\usepackage[latin1]{inputenc}
\usepackage{latexsym}
\usepackage{ifpdf}
\usepackage{pgf}
\usepackage{graphicx}
\usepackage{xypic}

\begin{document}

\title{From cosmic deceleration to acceleration: new constraints
from SN Ia and BAO/CMB}
\author{R. Giostri,}
\author{M.Vargas dos Santos,}
\author{I. Waga,}
\author{R. R. R. Reis,}
\author{M. O. Calv\~ao,}
\author{and B. L. Lago}
\affiliation{Instituto de F\'\i sica, Universidade
Federal do Rio de Janeiro\\{C. P. 68528, CEP 21941-972, Rio de Janeiro, RJ,
Brazil}}

\emailAdd{rgiostri@if.ufrj.br}
\emailAdd{vargas@if.ufrj.br}
\emailAdd{ioav@if.ufrj.br}
\emailAdd{ribamar@if.ufrj.br}
\emailAdd{orca@if.ufrj.br}
\emailAdd{brunolz@if.ufrj.br}

\abstract{
We use type Ia supernovae (SN Ia) data in combination with recent baryonic acoustic oscillations (BAO) and cosmic microwave background (CMB) observations to constrain a kink-like parametrization of the deceleration parameter ($q$). This $q$-parametrization can be written in terms of the initial ($q_i$) and present ($q_0$) values of the deceleration parameter, the redshift of the cosmic transition from deceleration to acceleration ($z_t$) and the redshift width of such transition ($\tau$). By assuming a flat space geometry, $q_i=1/2$ and adopting a likelihood approach to deal with the SN Ia data we obtain, at the $68\%$ confidence level (C.L.), that: $z_t=0.56^{+0.13}_{-0.10}$, $\tau=0.47^{+0.16}_{-0.20}$ and $q_0=-0.31^{+0.11}_{-0.11}$ when we combine BAO/CMB observations with SN Ia data processed with the MLCS2k2 light-curve fitter.  When in this combination we use the SALT2 fitter we get instead, at the same C.L.: $z_t=0.64^{+0.13}_{-0.07}$, $\tau=0.36^{+0.11}_{-0.17}$ and $q_0=-0.53^{+0.17}_{-0.13}$. 
Our results indicate, with a quite general and model independent approach, that MLCS2k2 favors Dvali-Gabadadze-Porrati-like cosmological models, while SALT2 favors $\Lambda$CDM-like ones. Progress in determining the transition redshift and/or the present value of the deceleration parameter depends crucially on solving the issue of the difference obtained when using these two light-curve fitters.}

\keywords{ dark energy theory, supernova type Ia - standard candles, baryon acoustic oscillations, cosmological parameters from CMBR}


\maketitle


\section{Introduction}

By the end of the last century,  type Ia supernovae (SN Ia) observations \cite{Riess1998} suggested that the expansion of the Universe is not slowing down, as would be expected from attractive gravity, but is speeding up. During the last years, in spite of several sources of systematics, that are still not completely well understood  \cite{Howell2010}, cosmic acceleration has been confirmed by subsequent SN Ia surveys (for recent compilations see \cite{Wood-Vasey2007,conley2011,kessler09,Amanullah2010}) and also by other probes, like, for instance, cosmic microwave background (CMB) and baryon acoustic oscillations (BAO) \cite{{percival10},blake11}, that in combination have corroborated the SN Ia result. Discovering the source of cosmic acceleration is one of the biggest challenges of modern cosmology. Fortunately, several experiments are underway, and others are being planned, that could bring a major progress in the determination of the cosmic expansion history, helping us to understand the fundamental physics behind cosmic acceleration. 

Many theoretical approaches to investigate cosmic acceleration have been suggested. Here we are interested in the so-called kinematical approach \cite{kinematic, ishida08} in which the deceleration parameter ($q$) is parametrized as a function of the redshift ($z$). Why $q(z)$ is considered and not another quantity like the equation of state parameter ($w$)? The main reason for this choice is that with the $q$-parametrization only a metric theory of gravity is assumed (one is not restricted to general relativity) and the assumptions on the dark sector are minimum. Of course we should keep in mind that by parametrizing $q$ (or any other cosmological quantity) one is not driven solely by the data. In fact, by defining how the deceleration parameter changes with redshift, a theoretical choice is performed. This choice should be considered a good one, if it is able to generalize, or at least describe, a large number of viable cosmological models at as large as possible a redshift range. By viable cosmology we mean one that has a kind of early matter dominated phase, in which structure might form, and a recent acceleration phase in accordance with observations. 

In this work we use recent SN Ia, BAO and CMB data to investigate the transition from cosmic deceleration to acceleration. We are particularly interested in the redshift of this transition, its (redshift) duration and the present value of the deceleration parameter. We remark that although the numerical values we shall obtain are conceivably sensitive to our specific $q(z)$ parameterization, differently from other $q$-parametrizations in the literature, our kink-like $q(z)$ expression is quite general and valid for a wide redshift range. As will be discussed in the next section, it generalizes several cosmological models and can be extended to high redshift. 

This paper is organized as follows: in section \ref{sec:kink}, we present our kink-like $q$-parametrization, discuss some of its properties and show how it generalizes known cosmological models. In section \ref{sec:obs}, we present the cosmological tests (SN Ia and BAO/CMB) we use to constrain the model parameters. In section \ref{results} our main results are exhibited and we conclude  in section \ref{sec:conc}.


\section{The kink-like $q(z)$ parametrization}
\label{sec:kink}
\subsection{Basic equations}
\label{sec:kink:basic}
In this work, we investigate flat cosmological models whose deceleration parameter, after radiation domination, can be described by the following expression \cite{ishida08} 
\begin{equation}
q(z)\equiv q_{f}+\frac{(q_{i}-q_{f})}{1-\frac{q_{i}}{q_{f}}\left( \frac{%
1+z_{t}}{1+z}\right) ^{1/\tau }}.  \label{novoq}
\end{equation}%
Here the parameter $z_{t}$ denotes the transition redshift ($q(z_{t})=0$)
from cosmic deceleration ($q>0$) to acceleration ($q<0$),  $q_{i}>0$  and $%
q_{f}<0$ are, respectively, the initial $(z\gg z_{t})$ and final $%
(z\rightarrow -1)$ asymptotic values of the deceleration parameter. The parameter $\tau$ is
associated with the width of the transition. It is related to the derivative of $q$ with respect to the redshift at $z=z_{t}$, that is to the jerk ($j$) (see eq.(\ref{jerk}) for definition) at the transition   
\begin{equation}
\tau ^{-1}=\left( \frac{1}{q_{i}}-\frac{1}{q_{f}}\right) \left[ \frac{dq(z)}{
d\ln (1+z)}\right] _{z=z_{t}} = \left( \frac{1}{q_{i}}-\frac{1}{q_{f}}\right) j(z_t).  \label{tau}
\end{equation}
In figure $1$ (upper panel) we plot $q(z)$ for $z_t=1$ and different values of the parameter $\tau$. In the lower panel we show $q(z)$ for $\tau= 1/3$ and different values of the parameter $z_t$. In both panels it has been assumed $q_i=1/2$ and $q_f=-1$.

\begin{figure*}[tbp]
\begin{center}
\includegraphics[scale=0.4]{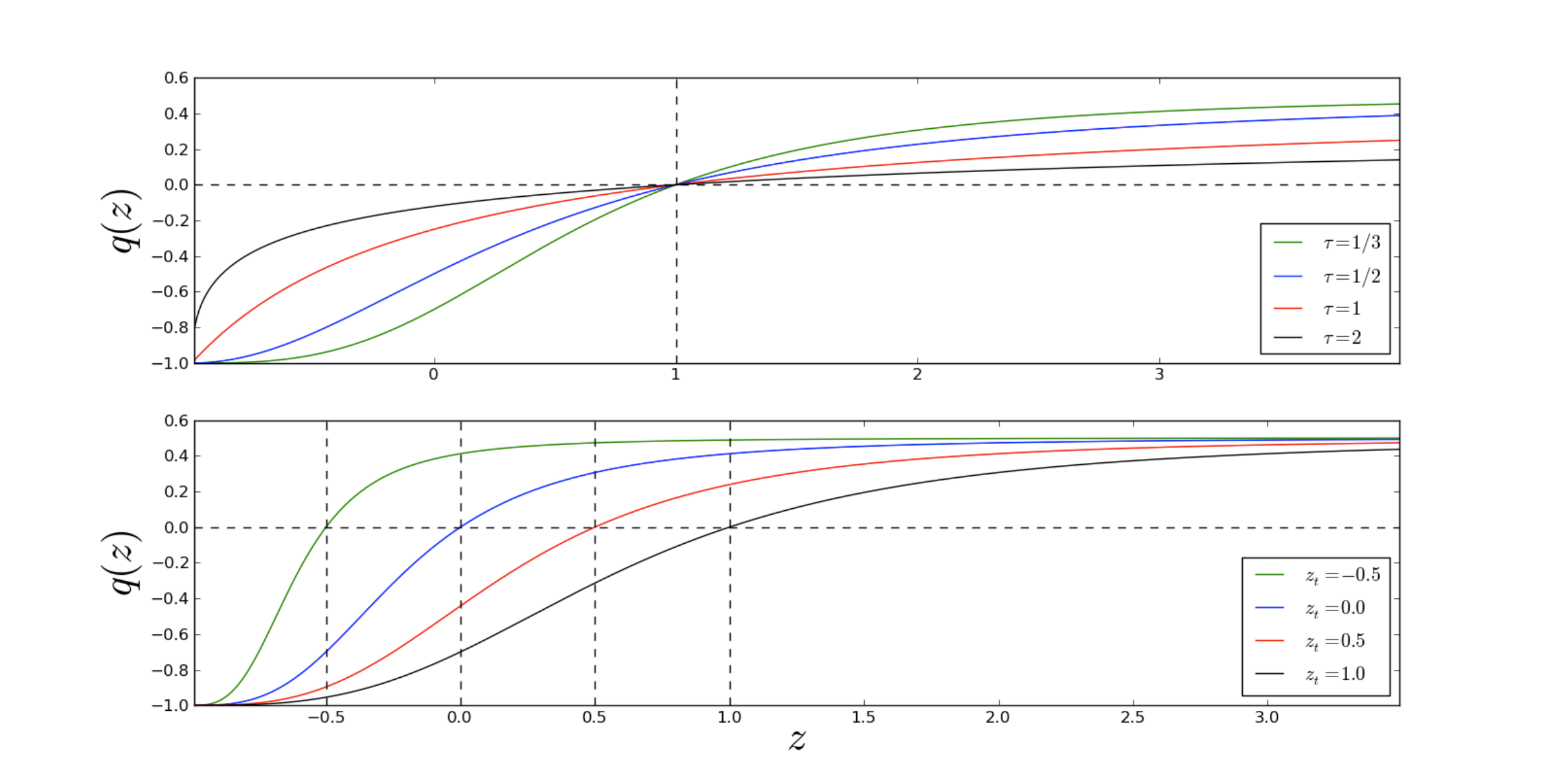}
\end{center}
\caption{\small{Dependence of the $q$-parametrization on the parameters. \textit{Top}: 
dependence on $\tau$ for $z_t=1$. \textit{Bottom}: dependence on $z_t$ for $\tau=1/3$. In both cases $q_i=0.5$ and $q_f=-1$.}}
\label{fig1}
\end{figure*}
 
From the definitions of the Hubble parameter, 
\begin{equation}
H:= \frac{\dot{a}}{a},
\end{equation}%
and the deceleration parameter
\begin{equation}
q:= -\frac{a\ddot{a}}{\dot{a}^{2}}=\frac{d}{dt}\left( 
\frac{1}{H}\right) -1=\frac{d\ln H}{d\ln (1+z)}-1,
\end{equation}%
we can write
\begin{equation}
H=H_{0}\exp \int_{x=0}^{z}\left( 1+q(x) \right) d\ln (1+x).  \label{hubblepar}
\end{equation}%
In the above equations, as usual, $a=1/(1+z)$ is the scale factor of the
Friedman-Robertson-Walker metric, $H_0=100h$ km/(s Mpc) is the present value of the Hubble parameter and a dot over a quantity denotes differentiation with respect to the cosmic time.  By using  eq. (\ref%
{novoq}),  eq. (\ref{hubblepar}) can be integrated to give
\begin{eqnarray}
\left( \frac{H(z)}{H_{0}}\right) ^{2}&=&\left( 1+z\right)
^{2(1+q_{i})} 
 \left( \frac{q_{i}\left( \frac{1+z_{t}}{1+z}\right) ^{1/\tau }-q_{f}
}{q_{i}\left( 1+z_{t}\right) ^{1/\tau }-q_{f}}\right) ^{2\tau (q_{i}-q_{f})} \notag \\ &=&\left( 1+z\right) ^{2(1+q_i)}  
 \left( \Omega _{m\infty }^{\frac{1}{2\tau (q_i-q_{f})}}+(1-\Omega
_{m\infty }^{\frac{1}{2\tau (q_i-q_{f})}})(1+z)^{-\frac{1}{\tau }}\right)
^{2\tau (q_i-q_{f})},
\label{hq0}
\end{eqnarray}%
where, in the last equality, we have introduced the quantity 
\begin{equation}
\Omega _{m\infty }=\left( 1-\frac{q_i}{q_{f}}\left( 1+z_{t}\right) ^{1/\tau
}\right) ^{2\tau (q_f-q_{i})},  \label{omeff}
\end{equation}%
defined as 
\begin{equation}
\Omega _{m\infty } := \lim_{z\rightarrow \infty }\left( \frac{H(z)}{H_{0}%
}\right) ^{2}\frac{1}{( 1+z) ^{2(1+q_i)}}. \label{omeff2}
\end{equation}%
The above definition is very convenient and facilitates the comparison of special cases of our $q$-parametrization with more traditional cosmological models like $\Lambda$CDM, $w$CDM, etc. At this point we stress that  the purpose of our $q$-parametrization, eq. (\ref{novoq}), is  to describe the late time transition from a cosmic decelerated to an accelerated phase. Although its validity can be extended to high redshift ($z\lesssim 1-2 \times10^3$), it does not describe the behavior of the deceleration parameter in the very early universe, that is, during the radiation dominated era (RDE) (when $q = 1$). Therefore, in eq. (\ref{omeff2}) the limit should be understood as $z>>z_t$, but after RDE. 

Another useful dimensionless quantity is the jerk \cite{chiba98}
\begin{equation}
j:= \frac{a^{2}\dddot{a}}{\dot{a}^{3}}= q\left(2q+1\right)-\frac{\dot{q}}{H}=q\left(2q+1\right)+\frac{dq}{d\ln (1+z)}.\label{jerk}
\end{equation}
With the $q$-parametrization given by eq. (\ref{novoq}), we obtain for the jerk the following expression
\begin{equation}
j(z)=\frac{q_f q_i \left(q_f (1+2 q_i) \tau +(1+2 q_f) q_i \left(\frac{1+z_t}{1+z}\right)^{2/\tau } \tau -\left(\frac{1+z_t}{1+z}\right)^{\frac{1}{\tau }} (-q_f+q_i+(q_f+q_i+4 q_f q_i) \tau )\right)}{\left(q_f-q_i \left(\frac{1+z_t}{1+z}\right)^{\frac{1}{\tau }}\right)^2 \tau }.\label{jerkq}
\end{equation}
Notice that $j(z_t)=\frac{q_f q_i}{\tau (q_f - q_i)}$, in accordance with eq. (\ref{tau}). In figure $2$ we plot $j(z)$ for $q_i=1/2$, $q_f=-1$,  $z_t=1$ and different values of $\tau$. Models with $\tau>1/3$ have $j(z)$ bellow one, while those with $\tau<1/3$, have $j(z)$ above the line $j=1$. Changing $z_t$ does not alter this feature; the only difference is the redshift position of the maximum ($\tau<1/3$) or the minimum ($\tau>1/3$). Higher $z_t$, higher the redshift of the $j(z)$ maximum/minimum.    

\begin{figure*}[tbp]
\begin{center}
\includegraphics[scale=0.4]{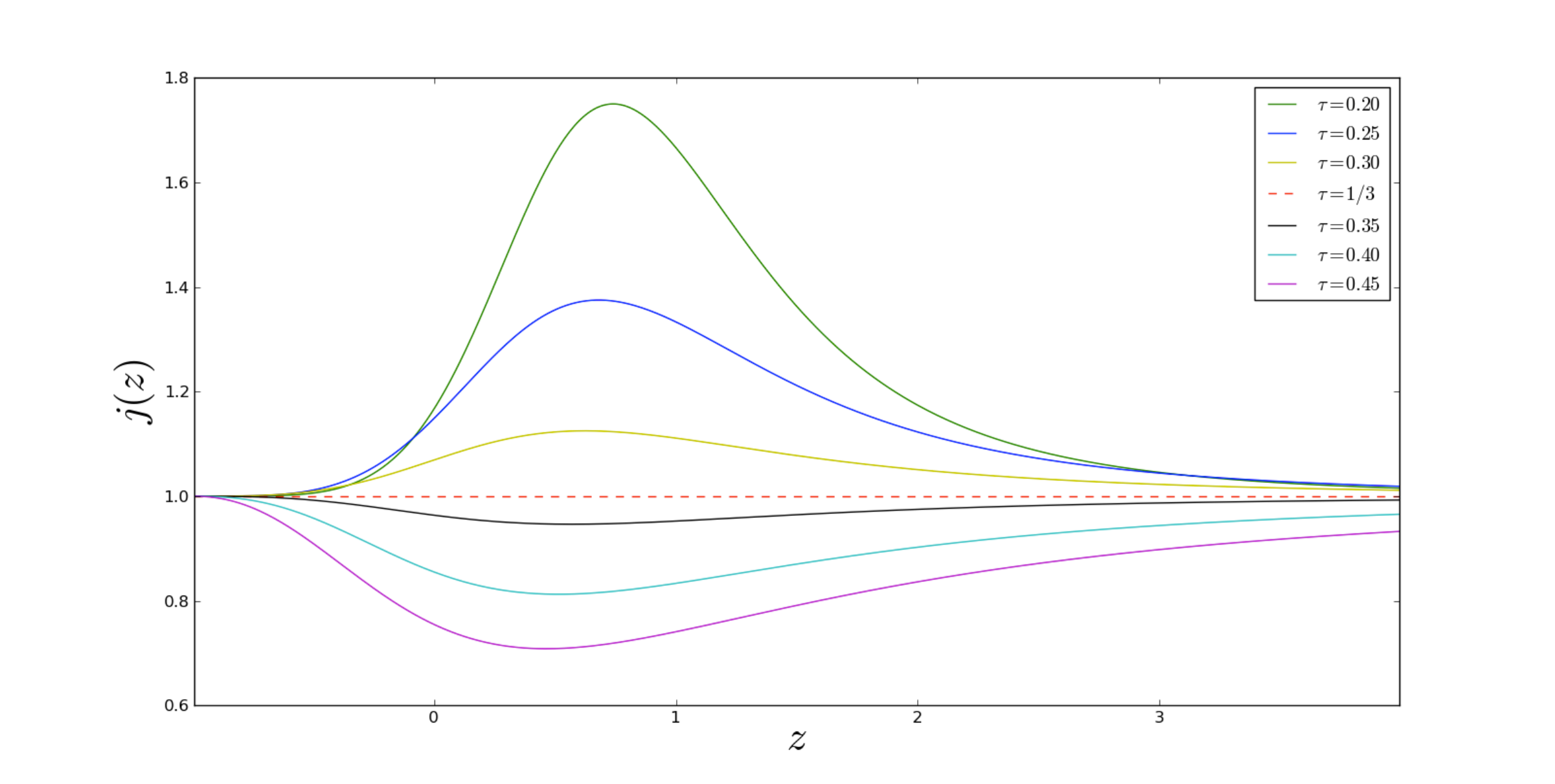}
\end{center}
\caption{\small{Dependence of the jerk $j(z)$ on $\tau$ for $z_t=1$, $q_i=0.5$ and $q_f=-1$.}}
\label{fig2}
\end{figure*}

From eq. (\ref{novoq}) we can relate the final value of the deceleration parameter, $q_f:= q(z=-1)$, to its current value $q_0:= q(z=0)$
\begin{equation}
q_f=\frac{q_i \left(1+z_t \right)^{1/\tau}}{1-\frac{q_i}{q_0}\left(1-(1+z_t)^{1/\tau}\right)}. \label{qfq0}
\end{equation}
By substituting eq. (\ref{qfq0}) in (\ref{novoq}) we get
\begin{equation}
q(z)=\frac{ q_0
   \left(\left(\frac{1+z_t}{1+z}\right)^{\frac{1}{\tau
   }}-1\right)}{(1+z)^{-1/\tau } \left(\frac{q_0}{
   q_i}+(1+z_t)^{\frac{1}{\tau }}-1\right)-\frac{q_0}{q_i}}.
   \label{newq}
\end{equation}
As observed above, with our parametrization we aim to describe the transition from a cosmic decelerated phase to an accelerated one. Therefore, since $q_i>0$ and $q_f<0$, by using eq. (\ref{qfq0}) it is straightforward to show that the parameter $\tau$ is constrained to the interval
\begin{equation}
0<\tau<\frac{\ln[1+z_t]}{\ln[1-\frac{q_0}{q_i}]}.
\label{tauconstr}
\end{equation}

\subsection{Special Cases }
\label{sec:kink:special}

In most scenarios, large scale structure formation requires that at early times the universe passes through a kind of matter dominated era in which, $H^2 \propto (1+z)^3$ and, consequently,  $q=1/2$. As discussed above, our $q$-parametrization is designed to describe cosmic evolution starting from a decelerated phase on and, for the sake of simplicity, we fix $q_i=1/2$ in this work. In fact, with this assumption we do not lose much generality, and are essentially excluding from our description models with a coupling in the dark sector (see reference \cite{ishida08} for more on this point). 

We present below some special cases that can be described by our parametrization.

\begin{enumerate}

\item The flat standard cosmological model with constant dark energy equation of state ($w$CDM)  ($w:=p/\rho=constant$, where $p$ is dark energy pressure and $\rho$ its energy density) is recovered if we identify  $\Omega_{m\infty }=\Omega _{m0}$ and $-3w=1/\tau =(1-2q_f)$. For $w$CDM the transition redshift is given by $z_t= \left[(\Omega_{m0}-1)(1+3w)/\Omega_{m0}\right]^{-1/(3w)}-1$. In particular, for flat cosmological constant + Cold Dark Matter cosmological model ($\Lambda$CDM) ($w=-1$), $q_{f}=-1$, $\tau =1/3$ and  $z_t=[2(1-\Omega_{m0})/$ $\Omega_{m0}]^{1/3}-1$. Note also that for $\Lambda$CDM $j=1$.

\item If we apply the definition of $\tau $, given by eq. (\ref{tau}), with $%
q_{i}=1/2$ and $q_{f}=-1$, to the flat DGP (DGP) brane-world model \cite{dvali00}
we obtain $\tau =1/2$, independent of $\Omega_{m0}$. Therefore $q$-models with 
$z_{t}=(2(1-\Omega_{m0})^{2}/$ $\Omega_{m0})^{1/3}-1$ (the DGP redshift
transition) and $\tau \approx 1/2$ are expected to be a good approximation
for DGP models.

\item Neglecting baryons the flat quartessence  Chaplygin model $(p=-M^{4(\alpha +1)}/\rho ^{\alpha
}) $ (see \cite{kamenshchik01}, \cite{bilic01}, \cite{bento02} and \cite{makler02}), is
obtained if (besides $q_i=1/2$) we assume $q_{f}=-1$, identify $1/\tau =3(1+\alpha
) $ and $\Omega _{m\infty }=(1-w_{0})^{1/(1+\alpha )}$, where $%
w_{0}=-(M^{4}/\rho _{0})^{\alpha +1}$ is the present value of the equation of
state parameter. For this model $z_t=\left(-2w_0/\left(1+w_0\right)\right)^\tau$.

\item Modified Polytropic Cardassian model \cite{gondolo03}, \cite{wang03}. This model depends on three
parameters: $m$ (denoted by $q$ in \cite{gondolo03}), $n$ and $\Omega _{m0}$. 
The kinematics of this model is obtained if we identify, $\Omega _{m0}=\Omega _{m\infty }$, $m=1/(\tau (1-2q_{f}))$ and $n=2/3(1+q_{f})$.
\end{enumerate}


\section{Observational Data}
\label{sec:obs}

In the next sections we derive the constraints set by three different experiments upon the parameters $z_t$, $\tau$ and $q_f$ (or $q_0$). We use the following datasets: a) observations of type Ia supernovae from the first-year SDSS-II Supernova Survey together with other supernova datasets (ESSENCE, SNLS, HST and a compilation of nearby SN Ia) as described in \cite{kessler09}; b) observations of baryon acoustic oscillations in the SDSS Luminous Red Galaxy sample \cite{percival10}, in the WiggleZ Survey \cite{blake11} and  6dF Galaxy Survey \cite{beutler11} datasets; c) measurements of cosmic microwave background temperature anisotropy from WMAP7 \cite{komatsu11, jarosik11}. In the following subsections,  the statistical analysis used for those observables is described.
 
\subsection{Type Ia supernovae}
\label{sec:obs:super}

In Ref. \cite{ishida08}, the $q$-parametrization given by eq. (\ref{novoq}) had been investigated by using two different SN Ia datasets: the \textit{Gold182} sample \cite{riess07}, processed with the light-curve fitter MLCS2k2 \cite{jha07},
and 115 SN Ia from the SNLS survey \cite{astier05}, analyzed with 
SALT \cite{guy05}. In this work we make use of only one SN Ia dataset, the 288 SN Ia compilation from Kessler \textit{et al.} \cite{kessler09} (sample ``e'' in their paper), and analyze it with both light-curve fitters MLCS2k2 \cite{jha07}  and SALT2 \cite{guy07}. In fact,  with SALT2 we use only 282 SN Ia, excluding the supernovae d086, 90O, 93B, 3256, 93O and Gilgamesh since they gave negative values for $\sigma_T^2$ (see eq. (\ref{sigmaT2}) for definition).

A key quantity in SN Ia investigation is the distance modulus
\begin{equation}
\mu_{th}(z;h,\boldsymbol{\theta}):= 5 \log\left[ D_L(z;\boldsymbol{\theta})\right] +\mu_0(h) . \label{eq:mu}
\end{equation}
Here $h$ is the Hubble constant in units of $100$ km/s Mpc$^{-1}$, $\boldsymbol{\theta}$ denotes the set of cosmological parameters of interest other than $h$, 
\begin{equation}
\mu_0(h):= 5\log\left(\frac{10^3 c/(\mbox{km/s})}{h}\right) = 42.38-5\log h.
\end{equation}
and
\begin{equation}
D_L(z;h,\boldsymbol{\theta})=(1+z)\int_0^z {\frac{dz'}{H(z';h,\boldsymbol{\theta})/H_0}}\;,
\end{equation}
is the dimensionless luminosity distance (in units of the Hubble distance today). In the following we discuss how to obtain constraints on cosmological parameters by using MLCS2k2 and SALT2 output as our data.

\subsubsection{\bf{MLCS2k2}}
In the MLCS2k2 fitting model \cite{jha07}, the variation among SN Ia light
curves is described with a single parameter ($\Delta$). Excess color variations
relative to the one-parameter model are assumed
to be the result of extinction by dust in the host galaxy
and in the Milky Way. The MLCS2k2 model magnitude, observed in
an arbitrary filter $Y$, is given by \cite{kessler09}
\begin{equation}
m_{Y,\epsilon}^{\rm model}=M_{Y',\epsilon}+p_{Y',\epsilon}\Delta
+q_{Y',\epsilon}\Delta^2
+K_{Y'Y,\epsilon}+\mu +X_{Y',\epsilon}^{\rm host} +X_{Y}^{\rm MW},
\label{mlcs2k2model}
\end{equation}
where $\epsilon$ is an epoch index that runs over the observations, $Y' \in \{U, B, V, R, I\}$ is one of the supernova rest-frame filters for 
which the model is defined, $\Delta$ is the MLCS2k2 shape-luminosity 
parameter that accounts for the correlation between peak luminosity
and the shape/duration of the light curve. Furthermore, the model for 
the host-galaxy extinction is 
$X^{\rm host}_{Y',\epsilon}=\zeta_{Y',\epsilon}(a_{Y'}+b_{Y'}/R_V)A_V$, 
where $\zeta_{Y',\epsilon}:=X^{\rm host}_{Y',\epsilon}/X^{\rm host}_{Y',0}$, 
and $a_{Y'}$, $b_{Y'}$ are constants; as usual, $A_V$ is the $V$ band extinction, at $B$ band peak 
($a_V=1$, $b_V=0$), and $R_V:=A_V/E(B-V)$, the ratio of $V$ band extinction
to color excess, at $B$ band peak. Finally, $X^{\rm MW}_{Y}$ is the
Milky Way extinction, $K_{Y'Y,\epsilon}$ is the $K$-correction between rest-frame and
observer-frame filters, and $\mu$ is the distance modulus.
The coefficients $M_{Y',\epsilon}$, $p_{Y',\epsilon}$, and $q_{Y',\epsilon}$
are model vectors that have been evaluated using nearly 100 well observed
low-redshift SN Ia as a training set.  Above, $\epsilon=0$ corresponds to the
$B$ band peak magnitude epoch.

Fitting the model to each SN Ia magnitudes, usually fixing $R_V$ 
gives $\mu$, $\Delta$, $A_V$ and $t_0$, the $B$-band peak magnitude
time.

In this work, to obtain the cosmological parameters constraints, we adopt the likelihood approach as described in \cite{lago11}. Assuming that all SN Ia events are independent, for the MLCS2k2 data, we can write the likelihood as
\begin{equation}
L(h,\boldsymbol{\theta},\sigma^{int})=\prod_{j=1}^N\frac{1}{\sigma_{T,j}\sqrt{2\pi}}
\exp\left[-\frac{1}{2}\chi^2_{\mathrm{MLCS2k2}}(h,\boldsymbol{\theta},\sigma^{int})\right], \label{likeMLCS2k2}
\end{equation}
where
\begin{equation}
\chi^2_{\mathrm{MLCS2k2}}(h,\boldsymbol{\theta},\sigma^{int}) := \sum_{j=1}^N \frac{[\mu_{j}-\mu_\mathrm{th}(z_j;h,\boldsymbol{\theta})]^2}{\sigma_{T,j}^2}. \label{eq:chi2_mlcs2k2}
\end{equation}
In the expression above $N$ is the total number of supernovae ($N=288$ for the MLCS2k2 analysis),  $\mu_{j}$ is the j-th SN Ia distance modulus ($\sigma^\mu_j$ is its statistical uncertainty, both estimated from the MLCS2k2 fitter), $z_j$ is the spectroscopically determined SN Ia redshift and 
\begin{equation}
\sigma_{T,j}^{2}=(\sigma^\mu_j)^2 +  (\sigma^\mu_{z_j})^2+(\sigma^{int})^2. \label{sigmaT}
\end{equation}
Following \cite{kessler09},  we assume an empty universe to project the redshift uncertainties onto distance modulus such that 
\begin{equation}
\sigma^\mu_{z_j}=\sigma^{z_j}\left(\frac{5}{\ln 10}\right)\frac{1+z_j}{z_j(1+z_j/2)}.\label{eq:z_prop}
\end{equation}
We checked that using an expression taking into account the cosmology has negligible effect on the constraints. Here, $\sigma^{z_j}$ is the $j$-th SN Ia total redshift uncertainty
\begin{equation}
(\sigma^{z_j})^2=(\sigma^{z_j}_{spec})^2+\sigma_{pec}^2,
\end{equation}
where $\sigma^{z_j}_{spec}$ is the error due the redshift determination and $\sigma_{pec}=0.0012$
is an additional error due to peculiar velocities. In eq. (\ref{sigmaT}), $\sigma^{int}$ is the intrinsic dispersion which is added in quadrature to the distance modulus and redshift dispersions. It represents intrinsic SN Ia dispersion (that are expected to exist even after corrections) and any potential systematic errors that have not been accounted for. In this work it is considered a free parameter to be determined from the data together with the other parameters (\cite{lago11}, see also \cite{kim11}). The approach we follow is different from the usual MLCS2k2 $\chi^2$ one in which $\sigma^{int}$ is estimated through an iterative procedure in which the $\chi^2$ per degree of freedom is made unity for the Hubble diagram constructed from the nearby SN Ia sample \cite{kessler09,sollerman09}. 

\subsubsection{\bf{SALT2}}
The aim of the SALT2 method \cite{guy07} is to model the mean evolution of the SN Ia spectral energy distribution (SED) and its variations with a few dominant components. It makes use of a two-dimensional surface in time and wavelength that describes the temporal evolution of the rest-frame SED. When using the SALT2 light-curve fitter, the SN Ia rest-frame flux at wavelength $\lambda$ is modeled by the following functional form 
\begin{equation}
\frac{dF_{rest}}{d\lambda}(t,\lambda)=x_0  \left[ M_0(t,\lambda) + x_1 M_1(t,\lambda)\right]  \exp [c\;  C(\lambda)]   \label{FSN}
\end{equation}
where $t$ is the rest-frame time since the date of maximum luminosity in B-band and $\lambda$ is the wavelength in the rest-frame of the SN Ia.  $M_0(t,\lambda)$ is the average spectral sequence whereas  $M_1(t,\lambda)$ is an additional component that describe the main variability of SN Ia (more additional components might be added). $C(\lambda)$ represents the average color correction law \cite{guy07}.  In contrast to $M_0$,  $M_1$ and $C$ that are properties of the global model (derived using a training set of SN Ia \cite{guy07}), $x_0$, $x_1$ and $c$ are parameters associated with each individual supernova. 

To compare with photometric SN Ia data, the observer frame flux in passband $Y$ is calculated as
\begin{equation}
F^Y_{obs}=(1+z)\int d\lambda'\left[\lambda'
\frac{dF_{rest}}{d\lambda'}(t,\lambda')T^Y(\lambda'/(1+z))\right],
\label{salt2obsflux}
\end{equation}
where $T^Y(\lambda)$ defines the transmission curve of the observer-frame filter $Y$, and $z$ is the redshift. Each SN Ia
light curve is fitted separately using eqs. (\ref{FSN}) and (\ref{salt2obsflux}) to determine the parameters $x_0$, $x_1$, and $c$ with corresponding errors. The parameter $c$ is a measure of the SN Ia color, $x_1$ is related to the stretch of the light curve and the peak rest-frame magnitude in the $B$ band is given by
\begin{equation}
m^{\ast}_B=-2.5\log [x_0\int d\lambda' M_0(t=0,\lambda')T^B(\lambda')]. 
\end{equation}
The $j$-th SN Ia distance modulus is modeled by
\begin{equation}
\mu_j = {m_B^*}_j - M + {\alpha} x_{1,j} -{\beta}   c_j \;,\label{eq:muSALT2}
\end{equation}
and  its statistical uncertainty by
\begin{equation}
(\sigma^\mu_j)^2=(\sigma^{m^*_B}_j)^2+{\alpha^2}(\sigma^{x_1}_j)^2 +{\beta^2} 
(\sigma^c_j)^2+2{\alpha }(\sigma^{m^*_B x_1}_j)-2{ \beta}(\sigma^{m^*_Bc}_j)-2{\alpha  \beta}(\sigma^{x_1c}_j)\;.\label{eq:sigmaSALT2}
\end{equation}
Here the quantities $(\sigma^{m^*_B}_j)^2$, $(\sigma^{x_1}_j)^2$, $(\sigma^c_j)^2$, $\sigma^{m^*_B x_1}_j$, $\sigma^{m^*_Bc}_j$ and $\sigma^{x_1c}_j$ are the components of the covariance matrix of $(m^*_B, x_1, c)$. We remark that $m_B^*$, $x_1$ and $c$ are derived from the fit to the light curves while the global parameters $\alpha$, $\beta$ and the absolute magnitude $M$ are estimated simultaneously with the cosmological parameters and are marginalized over when obtaining the cosmological parameter constraints.

Again assuming that all SN Ia events are independent, the likelihood for SALT2 data is given by
\begin{equation}
L(\boldsymbol{\theta},\alpha,\beta,\mathcal{M} (h,M),\sigma^{int})=\prod_{j=1}^{N}\frac{1}{\sigma_{T,j}\sqrt{2\pi}}
\exp\left[-\frac{1}{2}\chi^2_{\mathrm{SALT2}}(\boldsymbol{\theta},\alpha,\beta,\mathcal{M} (h,M),\sigma^{int})\right],
\label{likeSALT2}
\end{equation}
where
\begin{equation}
\chi^2_{SALT2}(\boldsymbol{\theta},\alpha,\beta,\mathcal{M}(h,M),\sigma^{int})=\sum\limits_{j=1}^{N}\frac{[\mu_j  -\mu_{th}(z_j;h,\boldsymbol{\theta}]^2}{\sigma_{T,j}^2}\;,\label{eq:chi2SALT2}
\end{equation}
$\mathcal{M}(h,M):=\mu_0(h)+M$, $ \mu_j $ is given by eq. (\ref{eq:muSALT2}),
\begin{equation}
\sigma_{T,j}^{2}=(\sigma^\mu_j)^2 +  (\sigma^\mu_{z_j})^2+(\sigma^{int})^2 \label{sigmaT2}
\end{equation}
and, for SALT2, $\sigma^\mu_j$ is given by eq. (\ref{eq:sigmaSALT2}) and $N=282$.
Alternatively, we can define the apparent magnitude in B band as the theoretical 
estimator for SALT2,
\begin{equation}
{m_B^*}_{th}(\boldsymbol{\xi_j};\boldsymbol{\theta},\alpha,\beta,\mathcal{M})=
 \mathcal{M} - {\alpha}  x_{1,j} +{\beta}  c_j +5 \log\left[ D_L(z_j;\boldsymbol{\theta})\right] 
\end{equation}
where $\boldsymbol{\xi_j}=(x_{1,j},c_j,z_j)$ denotes the set of observables and, for the considered model,
$\boldsymbol{\theta}=(q_i,q_f$ (or $q_0),z_t,\tau)$.

With the above definition we can rewrite (\ref{eq:chi2SALT2}) as
\begin{equation}
\chi^2_{SALT2} =\sum\limits_{j=1}^{N}\frac{[{m_B^*}_j - 
{m_B^*}_{th}(\boldsymbol{\xi_j};\boldsymbol{\theta},\alpha,\beta,\mathcal{M})]^2}{ (\sigma^\mu_j)^2 +  (\sigma^\mu_{z_j})^2+(\sigma^{int})^2}\;. \label{eq:chi2SALT2b}
\end{equation}

For the SALT2 data analysis we do not assume a fixed value for $\sigma^{int}$ \cite{sollerman09} nor obtain its value by imposing, through an iterative process, that $\chi^2$ per degree of freedom equals unity. As in the MLCS2k2 case, $\sigma^{int}$ is considered a free parameter and is determined from the data together with the other parameters \cite{lago11}. 
Although, in this work, our marginalization procedure is purely numerical, in the Appendix we discuss how to marginalize (\ref{likeSALT2}) analytically over the parameter $\mathcal{M}$ and (\ref{likeMLCS2k2}) over $\mu_0$. 


\subsection{Baryon acoustic oscillations and cosmic microwave background}
\label{sec:obs:BAO}

We start by defining the comoving sound horizon at the photon-decoupling epoch 
\begin{equation}
r_s(z_*)= \frac{c}{\sqrt{3}} \int_0^{1/(1+z_*)}\frac{da}{a^2H(a)\sqrt{1+(3\Omega_{b0} / 4 \Omega_{\gamma 0})a} },\label{eq:r_s}
\end{equation}
where $\Omega_{\gamma 0}$ and $\Omega_{b0}$ are, respectively, the present value of the photon and baryon density parameter. In eq. (\ref{eq:r_s}) $z_*$ is the redshift of photon decoupling and is well approximated by the formula given in \cite{hu96}. In accordance  with WMAP7 \cite{jarosik11}, in this work we use $z_* = 1091$ exactly. We checked that variations within the uncertainties about this value do not alter significantly the results  \cite{sollerman09}. Another relevant quantity is the redshift of the drag epoch ($z_d \approx 1020$), when the photon pressure is no longer able to avoid gravitational instability of the baryons. 

To obtain the BAO/CMB constraints we use the ``acoustic scale''
\begin{equation}
l_A=\pi\frac{d_A(z_*)}{r_s(z_*)}\quad,
\end{equation}
where  $d_A(z_*)= c \int_{0}^{z_*}dz'/H(z')$ is the comoving angular-diameter distance. Percival et al.  \cite{percival10} measured $r_s(z_d)/D_V(z)$, at $z=0.2$ and $z=0.35$.  Here,  
$D_V(z):=\left[ d_A^2(z) cz/H(z) \right]^{1/3}$ is the ``dilation scale'' introduced in \cite{eisenstein05}. In \cite{beutler11} the 6dF Galaxy Survey  also reported a new measurement of $r_s/D_V$ at $z=0.106$ and, more recently, the WiggleZ team \cite{blake11} obtained results at $z=0.44$, $z=0.60$ and $z=0.73$ (see table $1$).
Combining these results with the WMAP7-year \cite{jarosik11} value $l_A = 302.44\pm0.80$ give the values shown in table $1$ for  $\frac{d_A(z_*)}{D_V(z_{BAO})}\; \frac{r_s(z_d)}{r_s(z_*)}$. By using the WMAP7 \cite{jarosik11} recommended values for $r_s(z_d)$ and $r_s(z_*)$ we obtain $r_s(z_d)/r_s(z_*)=1.045\pm0.016$. By inserting this ratio in the expression above we obtain the BAO/CMB constraints ($\frac{d_A(z_*)}{D_V(z_{BAO})}$) also exhibited in table $1$.

\begin{table}
\begin{center}
\begin{tabular}{|@{} c@{} |@{}c @{}| @{}c @{} | @{}c@{} | @{} c @{}| @{} c@{}  | @{}c@{} |}
\hline 
\scriptsize{$z_{BAO}$} & \scriptsize{$0.106$} & \scriptsize{$0.2$}& \scriptsize{$0.35$}&\scriptsize{$0.44$}&\scriptsize{$0.6$}&\scriptsize{$0.73$}\\
\hline \hline
\scriptsize{$\frac{r_s(z_d)}{D_V(z_{BAO})}$} &\scriptsize{$ 0.336\pm0.015$\; }& \scriptsize{$0.1905\pm0.0061$\;}&\scriptsize{$ 0.1097\pm0.0036 \;$}&\scriptsize{$0.0916\pm0.0071\;$}&\scriptsize{$0.0726\pm0.0034\;$}& \scriptsize{$0.0592\pm0.0032\;$}\\
\hline
\scriptsize{ $\frac{d_A(z_*)}{D_V(z_{BAO})}\; \frac{r_s(z_d)}{r_s(z_*)}$} & \scriptsize{$ 32.35\pm1.45$ }& \scriptsize{$ 18.34\pm0.59$} & \scriptsize{ $ 10.56\pm0.35$} & \scriptsize{$  8.82\pm0.68$} & \scriptsize{$ 6.99\pm0.33$} & \scriptsize{ $ 5.70\pm0.31$}\\
\hline
\scriptsize{$\frac{d_A(z_*)}{D_V(z_{BAO})}$} &\scriptsize{$30.95\pm1.46$ }& \scriptsize{$17.55\pm0.60$ }&\scriptsize{$10.11\pm0.37$ }& \scriptsize{$ 8.44\pm0.67$}&\scriptsize{ $6.69\pm0.33$}& \scriptsize{$5.45\pm0.31$}\\
\hline
\end{tabular}
\caption{\small{Values of $\frac{r_s(z_d)}{D_V(z_{BAO})}$ \cite{percival10, beutler11, blake11}, $\frac{d_A(z_*)}{D_V(z_{BAO})}\; \frac{r_s(z_d)}{r_s(z_*)}$ and $\frac{d_A(z_*)}{D_V(z_{BAO})}$ for different values of $z_{BAO}$.}}
\end{center}

\end{table}

Now we can write the $\chi^2$ for the BAO/CMB analysis as
\begin{equation}
\chi^2_{BAO/CMB}={\bf X^tC^{-1}X},
\label{chi2baocmb}
\end{equation}
where
\begin{equation}
{\bf X}=\left(
          \begin{array}{cccc}
          \displaystyle\frac{d_A(z_*)}{D_V(0.106)} -30.95 \\
            \displaystyle\frac{d_A(z_*)}{D_V(0.2)} -17.55 \\
            \displaystyle\frac{d_A(z_*)}{D_V(0.35)} -10.11 \\
             \displaystyle\frac{d_A(z_*)}{D_V(0.44)} -8.44 \\
             \displaystyle\frac{d_A(z_*)}{D_V(0.6)} -6.69 \\
              \displaystyle\frac{d_A(z_*)}{D_V(0.73)} -5.45 \\
          \end{array}
        \right)
\end{equation}
and
\begin{equation}
{\bf C^{-1}}=\left(
          \begin{array}{cccccc}
          0.48435 & -0.101383 &-0.164945 &-0.0305703 &-0.097874 & -0.106738\\
           -0.101383 & 3.2882 & -2.45497 & -0.0787898 & -0.252254 & -0.2751\\
           -0.164945 & -2.45497 & 9.55916 & -0.128187 & -0.410404 & -0.447574\\
           -0.0305703 & -0.0787898 & -0.128187 & 2.78728 & -2.75632 & 1.16437\\
          -0.097874 & -0.252254 & -0.410404 & -2.75632 & 14.9245 & -7.32441 \\
         -0.106738 & -0.2751 & -0.447574 & 1.16437 & -7.32441 & 14.5022  \\
          \end{array}
        \right)
\end{equation}
is the inverse covariance matrix derived by using the above results together with the correlation coefficients $r=0.337$, $r=0.369$ and $r=0.438$ calculated for the  $r_s/D_V$  pair of measurements at $z=(0.2, 0.35)$, $z=(0.44, 0.6)$ and $z=(0.6, 0.73)$, respectively \cite{percival10,blake11}.


\section{Results}
 \label{results}
 
Through all this work, to obtain the probability distributions (PDFs), the Metropolis-Hasting algorithm has been used \cite{mcmc}. Generally, to obtain the PDFs,  5000 chains were generated with 10000 points for each chain. We first consider models in which the final value of the deceleration parameter is fixed to $q_f = -1$. Cosmological models in this special class asymptotically  tend to a de Sitter phase in the future. Several popular cosmological models like flat $\Lambda$CDM, quartessence Chaplygin, DGP etc, belong to this class as discussed in section \ref{sec:kink:special}. 
 
\begin{figure*}[tbp]
\begin{center}
\includegraphics[scale=0.5]{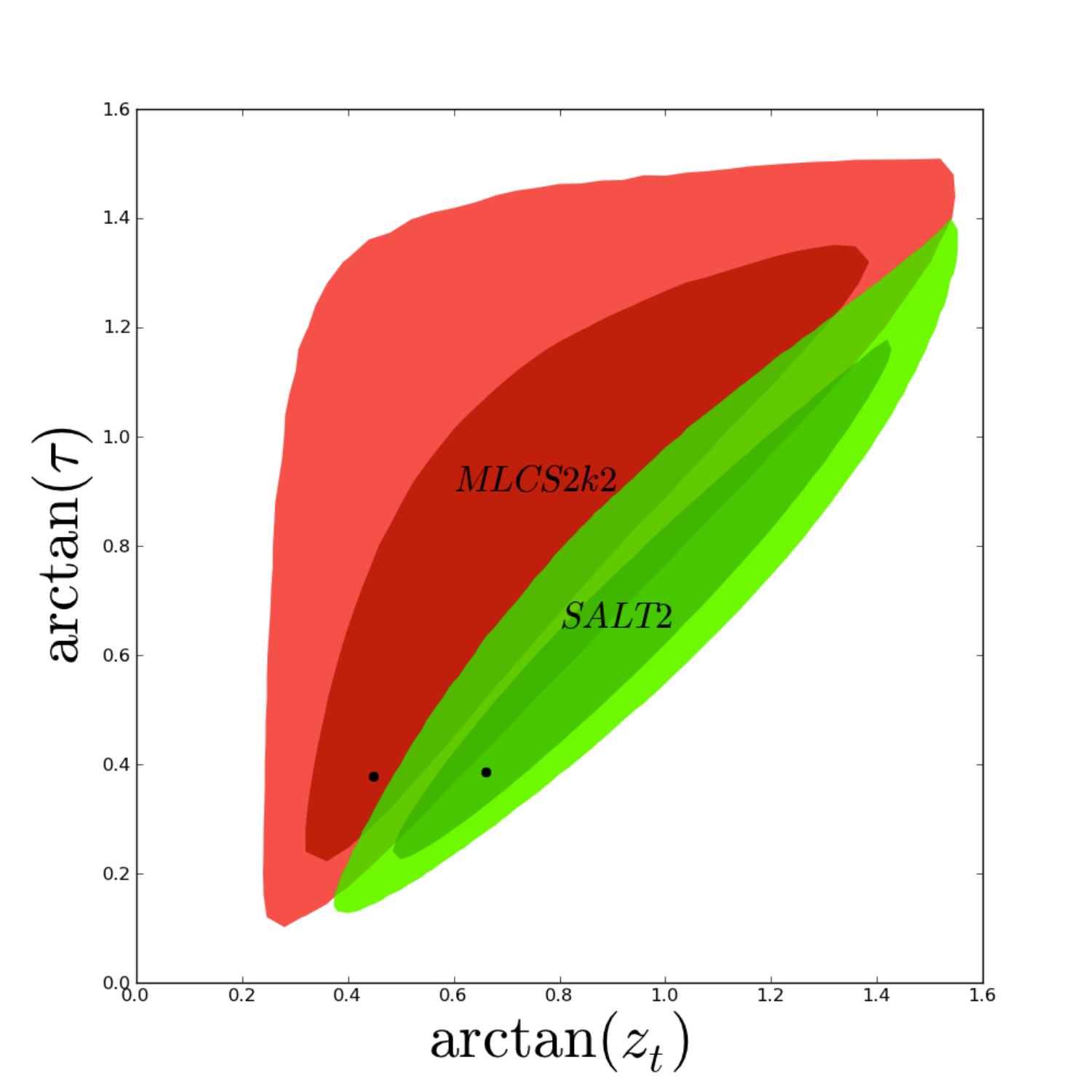}
\end{center}
\caption{\small{68\% and 95\% confidence contours in the plane $\arctan(z_t)$ vs. $\arctan(\tau)$, for
$q_i=0.5$ and $q_f=-1$, for SN Ia data only, for both MLCS2k2 (in red) and SALT2 (in green).
 We marginalized over
all other parameters with flat priors. 
The dots indicate the best-fit values.}}
\label{fig3}
\end{figure*}

By fixing $q_f=-1$, we are left with only two free model parameters, the transition redshift $z_t$ and its duration $\tau$.   In figure \ref{fig3} we display $68\%$ and $95\%$ confidence level (C.L.) regions in the $(\arctan{z_t}, \arctan{\tau})$ plane allowed by supernovae experiments. As mentioned in section \ref{sec:obs}, in this work we are considering the Kessler \textit{et al.} \cite{kessler09} SN Ia compilation. From figure \ref{fig3} we can see that, at high confidence level,  $z_t\leqslant0$ is not permitted, indicating that, from SN Ia observations, a transition occurred in the past. Furthermore it is also clear from the figure that, SN Ia observations alone are not able to strongly constrain the parameters $z_t$ and $\tau$. For instance, a high value of the transition redshift (let say $z_t \gtrsim 5$) is allowed by SN Ia observations if the transition is slow ($\tau \gtrsim 1$). Since SN Ia observations probe the universe only up to $z\sim 1-2$, the distance to an object, say, at $z\lesssim1$ in a universe with $\tau \gtrsim 1$ and high $z_t$, can be similar to the distance to the same object,  in a universe with $z_t\lesssim 1$ but with a faster transition (smaller $\tau$) and this explains the shape of the contours.

\begin{figure*}[tbp]
\begin{center}
\xymatrix{\includegraphics[scale=0.4]{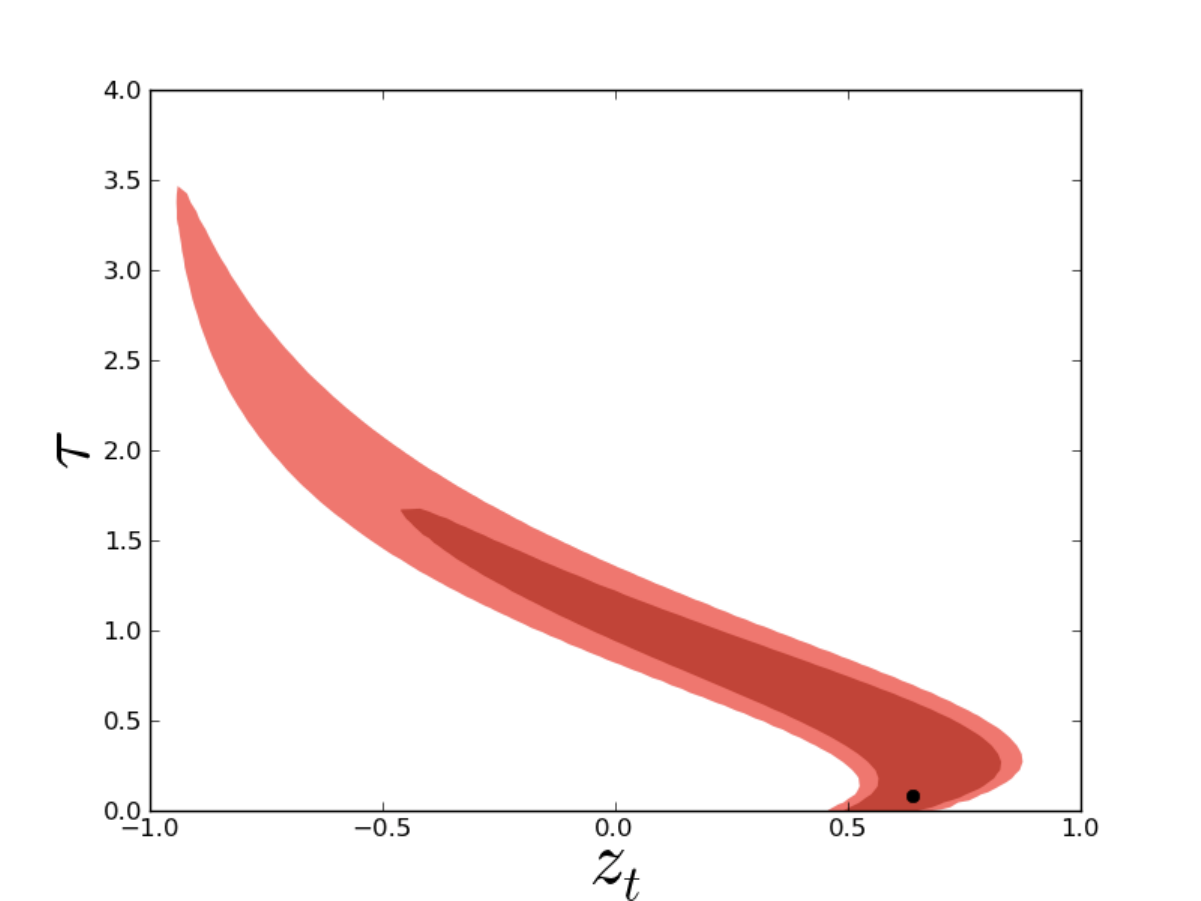}&%
\includegraphics[scale=0.4]{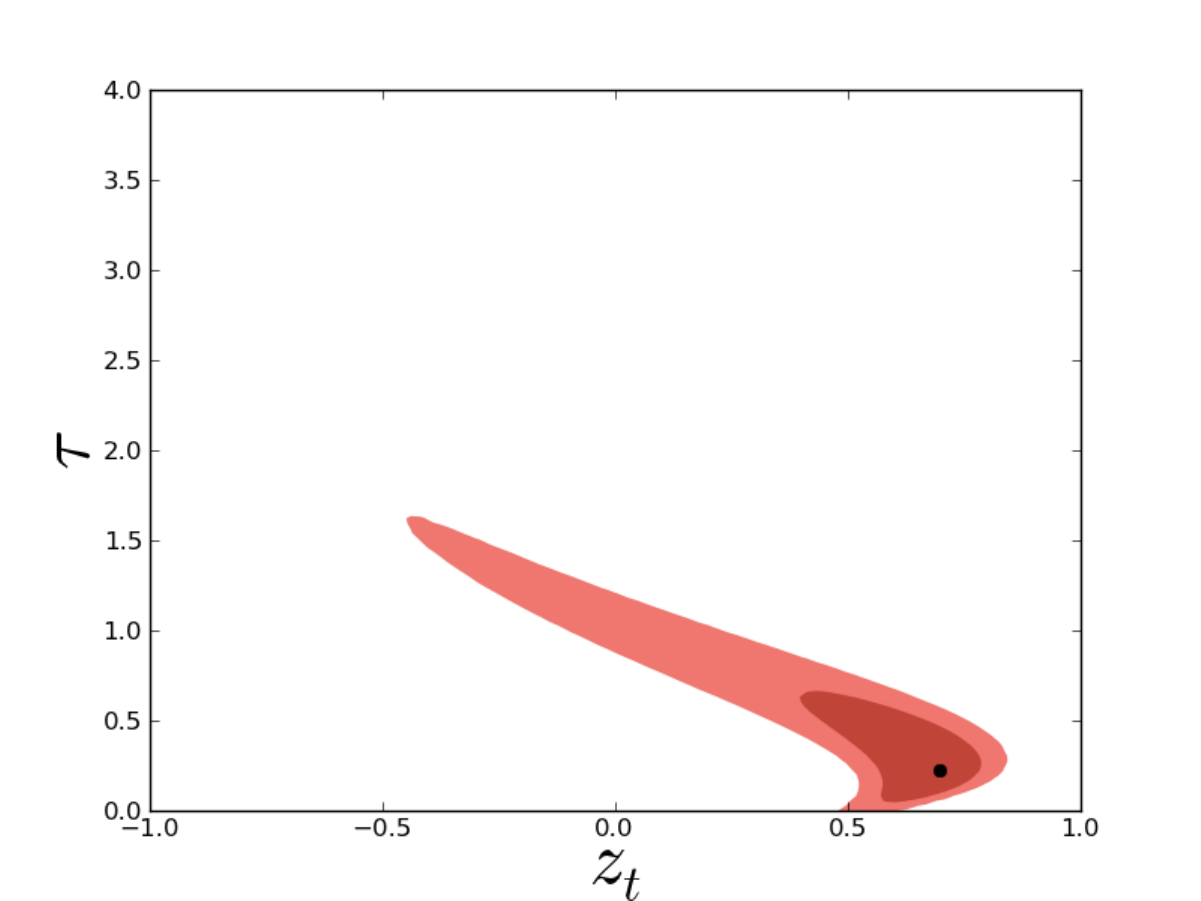}}
\end{center}
\caption{\small{68\% and 95\% confidence contours in the plane $z_t$ vs. $\tau$, for
$q_i=0.5$ and $q_f=-1$, for BAO/CMB data only. We marginalized over all other parameters
 with flat priors. \textit{Left}: result obtained
with only the two measurements from \cite{percival10}. \textit{Right}: result
obtained with all six measurements considered in this work. 
The dots indicate the best-fit values.}}
\label{fig4}
\end{figure*}

In figure \ref{fig4} we show BAO/CMB constraints on $z_t$ and $\tau$ in the case $q_f=-1$. In the left panel we display $68\%$ and $95\%$ confidence regions obtained using only the two measurements of $r_s(z_d)/D_V(z)$ obtained by Percival \textit{et al.} \cite{percival10}. In the right panel we display the results when six measurements of $r_s(z_d)/D_V(z)$)\cite{percival10,beutler11,blake11} are taken into account. The improvement these
new observations bring in constraining the model parameters is
evident. Notice also that this test is complementary to the SN Ia one and by combining them we can considerably reduce the allowed parameter space. 

\begin{figure*}[tbp]
\begin{center}
\includegraphics[scale=0.4]{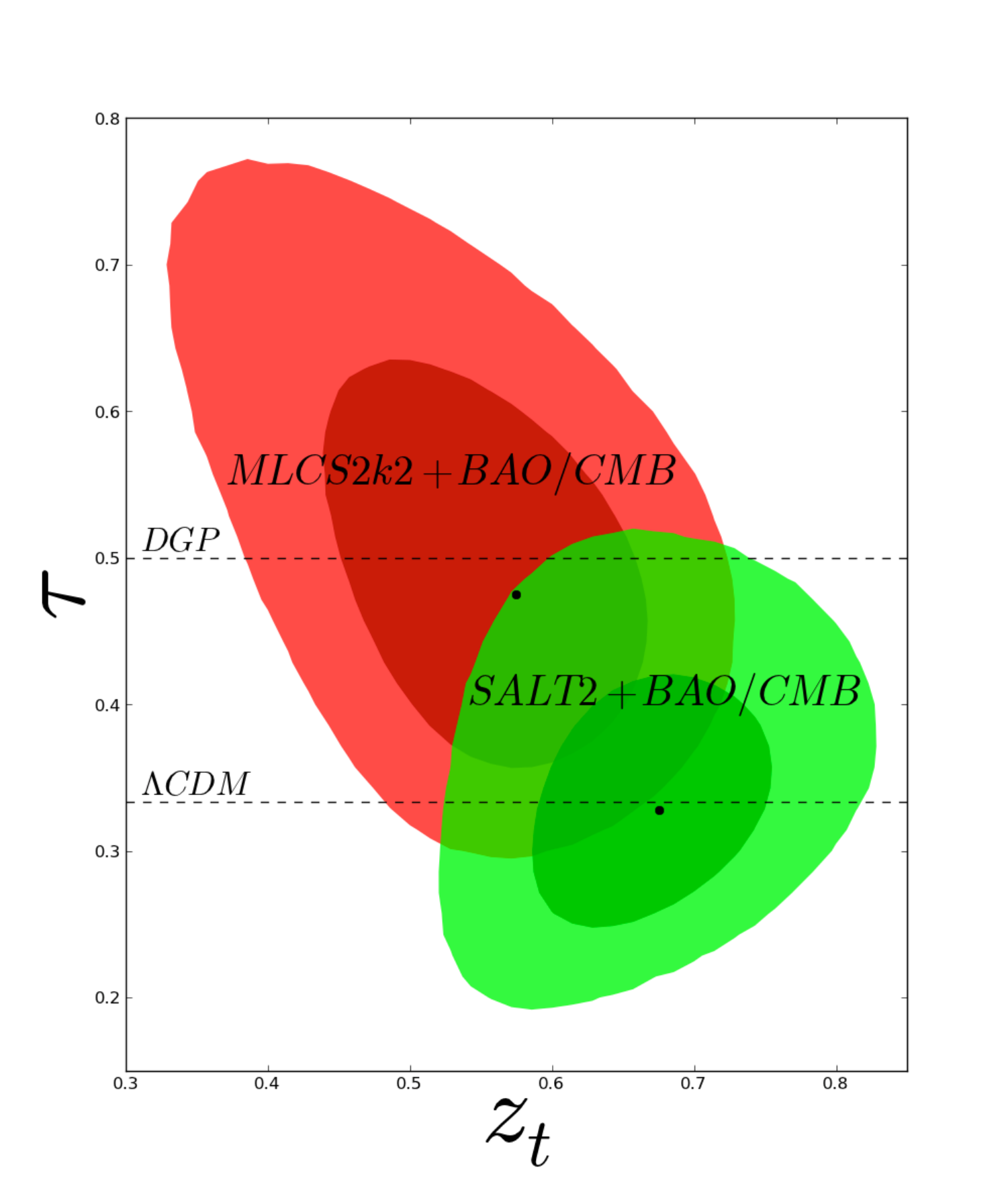}
\end{center}
\caption{\small{68\% and 95\% confidence contours in the plane $z_t$ vs. $\tau$, for
$q_i=0.5$ and $q_f=-1$, for SN Ia (MLCS2k2 in red and SALT2 in green) combined with BAO/CMB (six
measurements). We marginalized over all other parameters with flat priors. 
The dashed lines denote the values of the parameters that correspond
to the models $\Lambda$CDM ($\tau=1/3$) and DGP ($\tau=1/2$). 
The dots indicate the best-fit values.}}
\label{fig5}
\end{figure*}

Since the SN Ia data are independent from BAO/CMB we can write the combined $\chi^2$ statistics as
\begin{equation}
   \chi^2 = -2 \ln L_{SN} + \chi^2_{BAO/CMB}\;,
   \label{eq:chi2}
\end{equation}
where $ \chi^2_{BAO/CMB}$ is given by eq. (\ref{chi2baocmb}) and $L_{SN}$ by eq. (\ref{likeMLCS2k2}), when adopting MLCS2k2, or eq. (\ref{likeSALT2}) when the SALT2 light-curve fitter is used. In figure \ref{fig5} we show the $68\%$ and $95\%$ C.L. regions in the $(z_t, \tau)$ plane imposed by the combined data sets. In the figure the dotted lines represent the $\Lambda$CDM ($\tau=1/3$) and the DGP ($\tau=0.5$) models.  The figure indicates that the combination of the BAO/CMB data set with SN Ia analyzed with  MLCS2k2 tend to favor the DGP model, while the combination of BAO/CMB plus SN Ia analyzed with SALT2, tends to favor  the $\Lambda$CDM model.

\begin{figure*}[tbp]
\begin{center}
\includegraphics[scale=0.45]{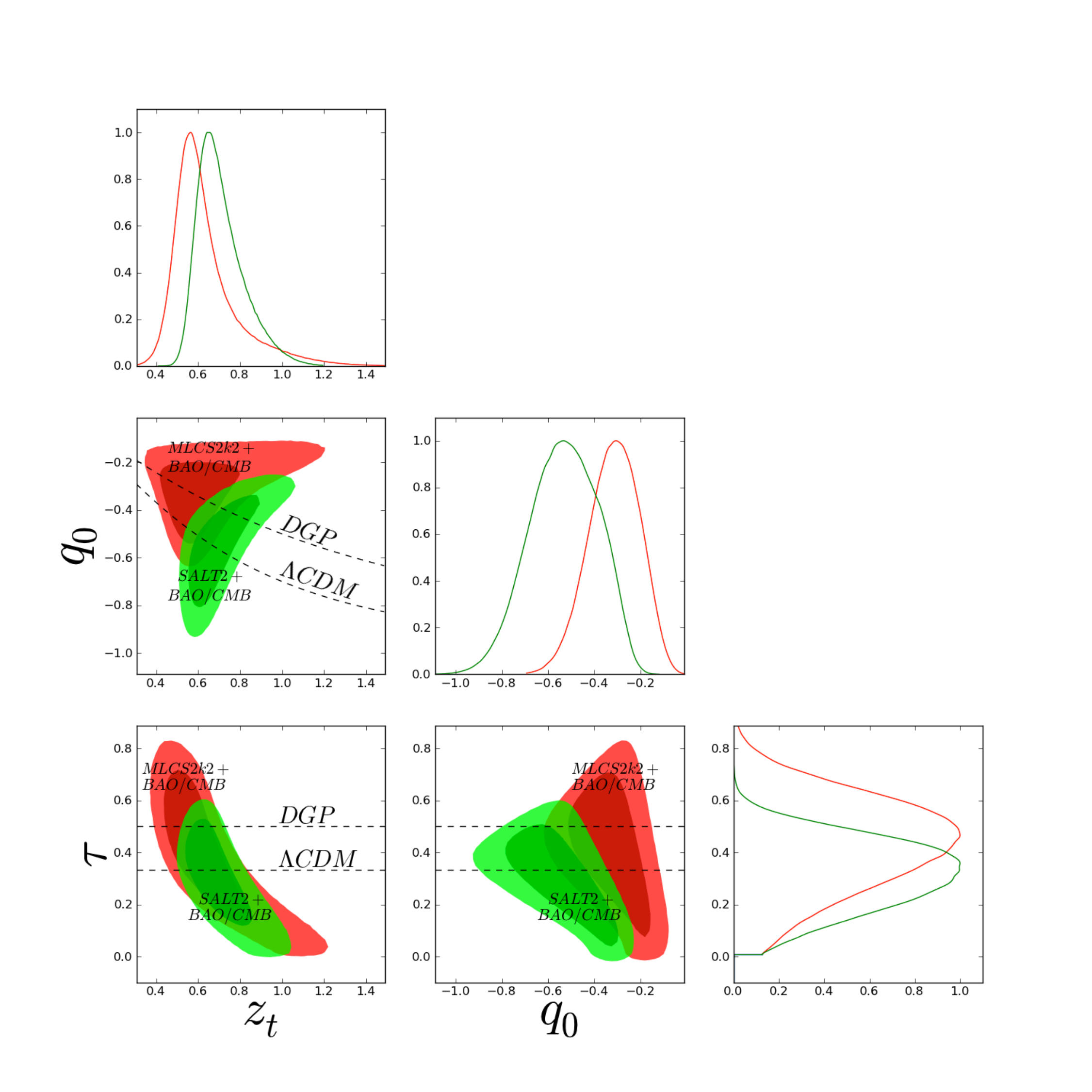}
\end{center}
\caption{\small{68\% and 95\% confidence contours in the planes $z_t$ vs. $\tau$ 
(bottom left), $z_t$ vs. $q_0$ (middle left) and $q_0$ vs. $\tau$ (bottom center), considering
$q_i=0.5$, for SN Ia (MLCS2k2 in red and SALT2 in green) combined with BAO/CMB (six
measurements). The dashed lines denote the values of the parameters that correspond
to the models $\Lambda$CDM ($\tau=1/3$) and DGP ($\tau=1/2$). 
We also show the 1D PDFs ($\mathcal{L}/\mathcal{L}_{max}$)
for $z_t$ (top), $q_0$ (middle center) and $\tau$ (bottom right). In all
cases we marginalized over all other parameters with flat priors.}}
\label{fig6}
\end{figure*}

We now consider the situation in which we do not impose a fixed value for $q_f$. In this quite general case, we would have to determine three model parameters, namely $z_t$, $\tau$ and $q_f$ (we still assume $q_i=1/2$). However, instead of working with $q_f$, we find more convenient to use $q_0$, the present value of the deceleration parameter.  By using eq. (\ref{qfq0}), one can re-express $q(z)$ in terms of $q_0$ (see eq. (\ref{newq})). We remark that possible $\tau$ values are not arbitrary but constrained by eq. (\ref{tauconstr}). 

\begin{figure*}[tbp]
\begin{center}
\includegraphics[scale=0.6]{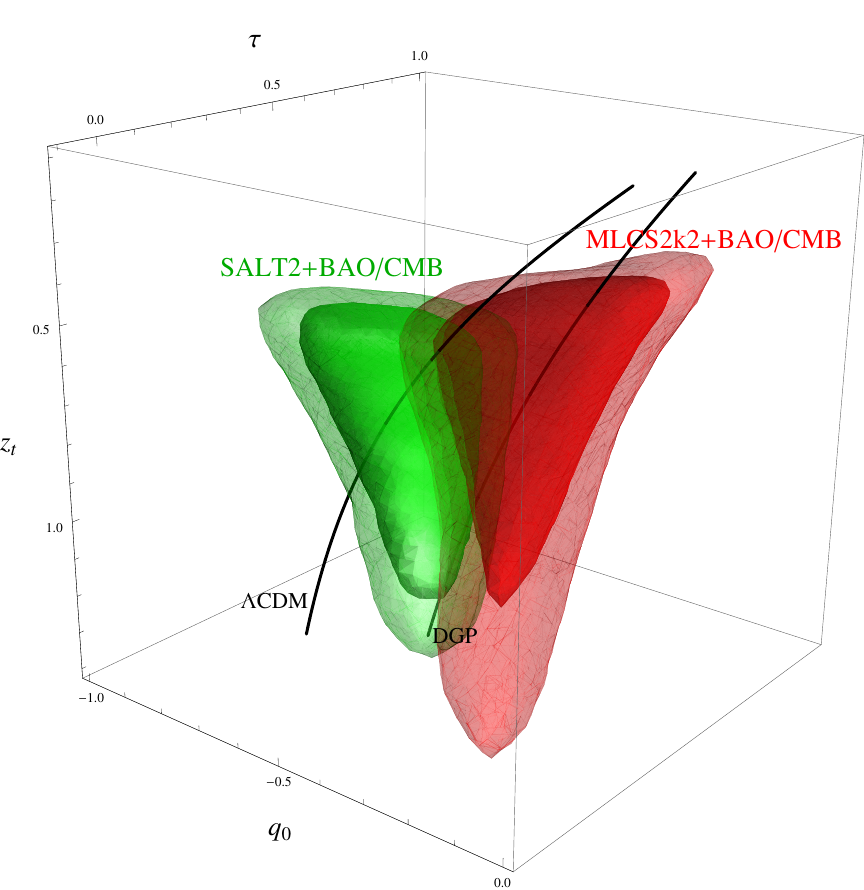}
\end{center}
\caption{\small{68\% and 95\% confidence surfaces in the space ($q_0$, $\tau$, $z_t$), 
for $q_i=0.5$, for SN Ia (MLCS2k2 in red and SALT2 in green) combined with BAO/CMB (six
measurements). We marginalized over all other parameters with flat priors. The lines represent
the flat models $\Lambda$CDM and DGP.}}
\label{fig7}
\end{figure*}

\begin{figure*}[tbp]
\begin{center}
\xymatrix{\includegraphics[scale=0.4]{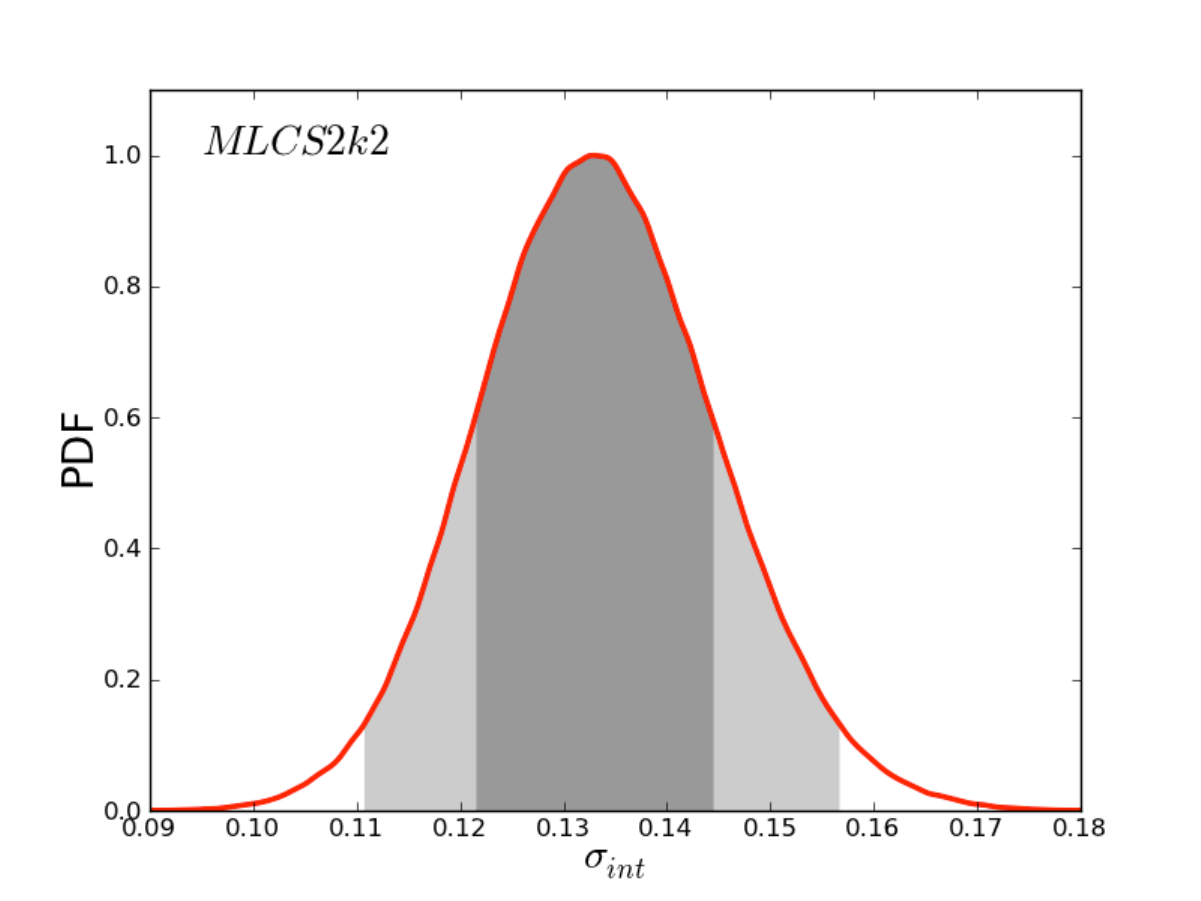}&%
\includegraphics[scale=0.4]{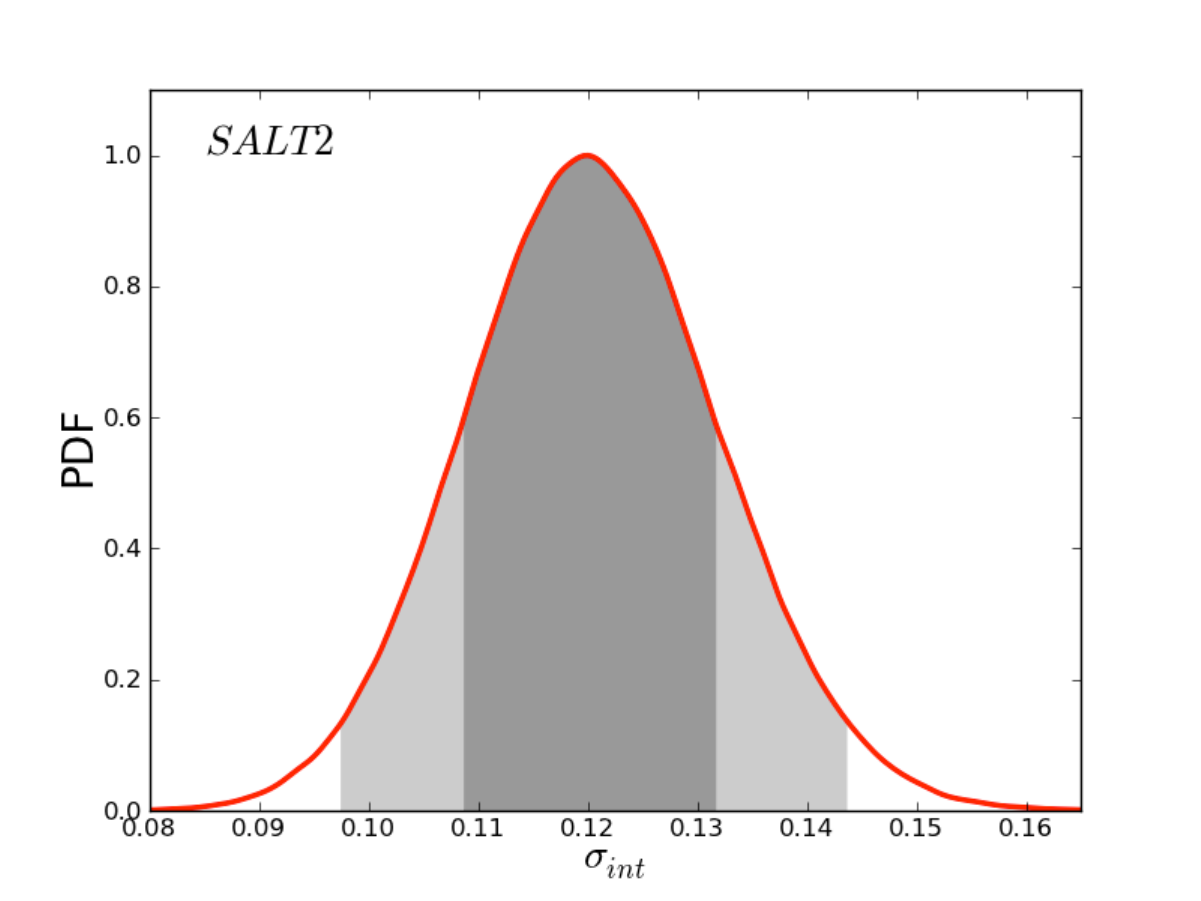}}
\end{center}
\caption{\small{1D probability distributions ($\mathcal{L}/\mathcal{L}_{max}$) for $\sigma_{int}$, marginalized over all other
parameters with flat priors.
The shaded regions indicate the 68\% and 95\% confidence intervals.}}
\label{fig8}
\end{figure*}

\begin{table}
\begin{center}
\begin{tabular}{| l | l | l | l |}
\hline 
&\multicolumn{1}{|c|}{MLCS2k2+BAO/CMB} & \multicolumn{1}{|c|}{SALT2+BAO/CMB} \\
\hline \hline
\multicolumn{1}{|c|} {$z_t$} &\multicolumn{1}{|c|} {$0.56 ^{+0.13 \;(+0.45)}_{-0.10 \;(-0.18)}$ }& \multicolumn{1}{|c|}{$0.64 ^{+0.13 \;(+0.30)}_{-0.07 \;(-0.12)}$}\\
\hline
 \multicolumn{1}{|c|}{$\tau$} &\multicolumn{1}{|c|}{$ 0.47^{+0.16\; (+0.27)}_{-0.20\; (-0.40)}$ }&\multicolumn{1}{|c|} {$ 0.36^{+0.11\; (+0.19)}_{-0.17\; (-0.30)}$}\\
\hline
 \multicolumn{1}{|c|}{$q_0$} &\multicolumn{1}{|c|}{$ -0.31^{+0.11\; (+0.20)}_{-0.11\; (-0.23)}$ }&\multicolumn{1}{|c|}{ $ -0.53^{+0.17\; (+0.28)}_{-0.13\; (-0.27)}$}\\
\hline
 \multicolumn{1}{|c|}{$\sigma_{int}$} &\multicolumn{1}{|c|}{$ 0.13^{+0.01\; (+0.02)}_{-0.01\; (-0.02)}$} &\multicolumn{1}{|c|}{ $ 0.12^{+0.01\; (+0.02)}_{-0.01\; (-0.02)}$}\\
\hline
\multicolumn{1}{|c|}{$\alpha$} &\multicolumn{1}{|c|} {-} &  \multicolumn{1}{|c|}{$0.094^{+0.011\; (+0.022)}_{-0.011\; (-0.022)}$}\\
\hline
\multicolumn{1}{|c|}{$\beta$} & \multicolumn{1}{|c|} {-} &\multicolumn{1}{|c|}{ $ 2.18^{+0.10\; (+0.21)}_{-0.11\; (-0.21)}$}\\
\hline
\end{tabular}
\caption{\small{Summary of the best-fit values for all parameters when using SN Ia + BAO/CMB,
including the 68\% and 95\% confidence intervals. For each parameter we marginalized over all other parameters with flat priors.}}
\end{center}
\end{table}

\begin{figure*}[tbp]
\begin{center}
\includegraphics[scale=0.5]{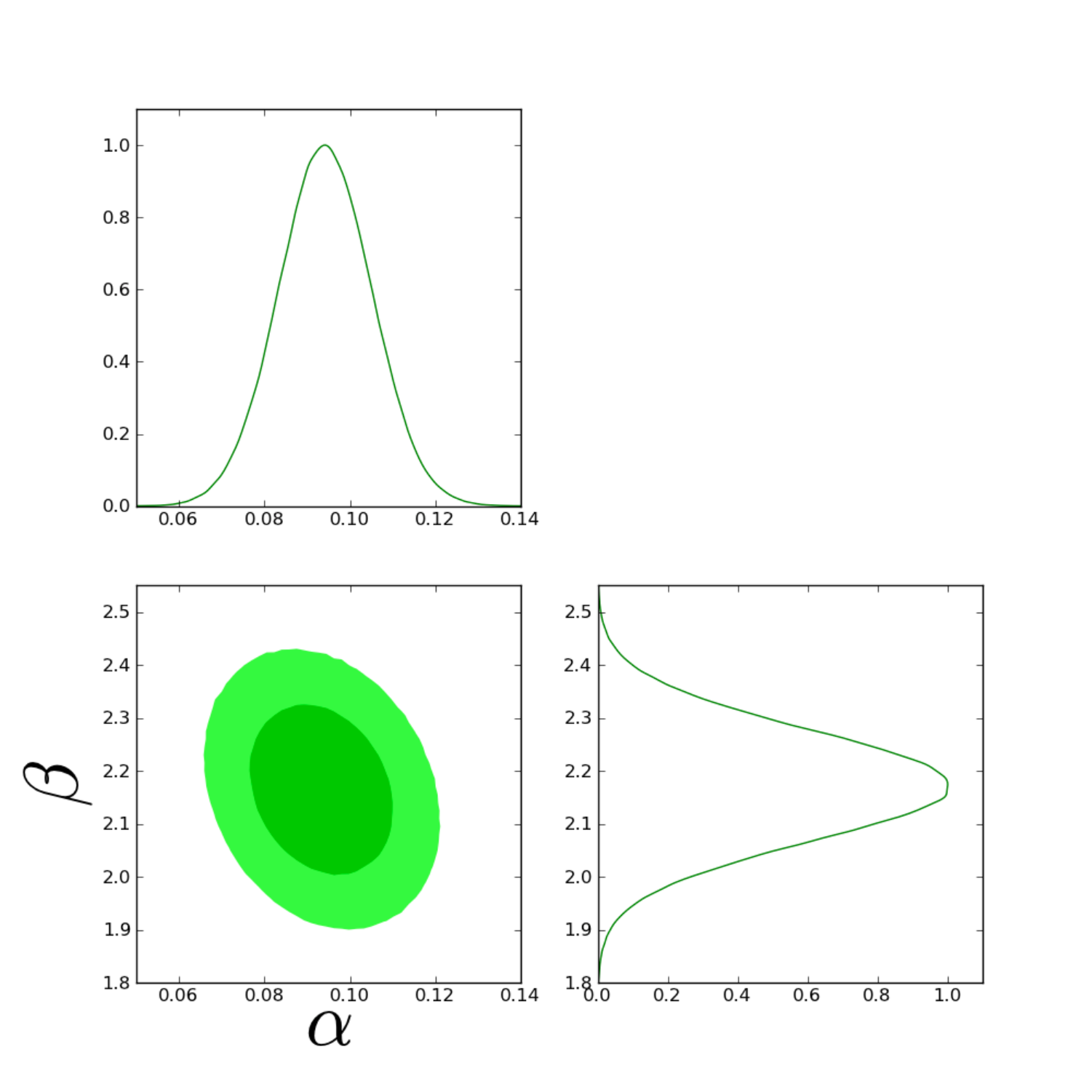}
\end{center}
\caption{\textit{Bottom left}: 68\% and 95\% confidence contours in the plane 
$\alpha$ vs. $\beta$, considering
SALT2 + BAO/CMB and marginalizing over all other parameters wiht flat priors.
\textit{Top}: 1D probability distribution ($\mathcal{L}/\mathcal{L}_{max}$) for $\alpha$. \textit{Bottom right}:
1D probability distribution ($\mathcal{L}/\mathcal{L}_{max}$) for $\beta$.}
\label{fig9}
\end{figure*}

In figure \ref{fig6} we display 2D marginalized (flat prior) C.L. regions ($68\%$ and $95\%$) in the $(z_t$, $\tau)$, $(z_t$, $q_0)$ and $(q_0$, $z_t)$ planes imposed by the combination of SN Ia (using MLCS2k2 and SALT2 fitters) and BAO/CMB tests. The marginalized (flat prior) 1D likelihoods
(normalized to its maximum values -- $\mathcal{L}/\mathcal{L}_{max}$) for each cosmological parameter, $z_t$, $\tau$ and $q_0$, are also shown.  In the figure are also displayed dashed lines corresponding to the flat DGP ($\tau=1/2$) and $\Lambda$CDM ($\tau=1/3$) cosmological models. In table 2 we show the best-fit and the $68\%$ and $95\%$ limits on several quantities after marginalizing over all the other parameters with a flat prior.  In figure \ref{fig7} the $68\%$ and $95\%$ C.L. 3D regions for the parameters $q_0$, $z_t$ and $\tau$ are shown. The curves corresponding to the flat $\Lambda$CDM and DGP models are also displayed. It is clear from the figure that the flat $\Lambda$CDM model is favored when we combine BAO/CMB with SN Ia analyzed with SALT2, while flat DGP is preferred if we use the MLCS2k2 fitter. In figure \ref{fig8} we show the 1D probability distribution, $\mathcal{L}/\mathcal{L}_{max}$ (marginalized with a flat prior), for the the intrinsic dispersion $\sigma_{int}$  and in figure \ref{fig9} the $68\%$ and $95\%$ C.L. regions in the $(\alpha, \beta)$ plane are displayed together with 1D distribution ($\mathcal{L}/\mathcal{L}_{max}$) for these parameters. 

\section{Discussion}
\label{sec:conc}

In this work we considered recent SN Ia, BAO and CMB observations to investigate the transition from cosmic deceleration to acceleration. Instead of adopting the traditional $\chi^2$ treatment, the SN Ia analysis was performed with the likelihood method, as described in \cite{lago11}. Following \cite{ishida08} we considered a kinematical approach using a kink-like parametrization for the deceleration parameter. We first considered the simple case in which the models have a final de Sitter phase such that $q_f=-1$.  Since \cite{ishida08} focus on this case, it is worthwhile to compare the results obtained in the present work with those in that paper. First let us remind that in \cite{ishida08} two different SN Ia datasets were used: the \textit{Gold182} sample \cite{riess07}, processed with the light-curve fitter MLCS2k2, and the SNLS survey \cite{astier05}, analyzed with SALT. Here we make use of only one SN Ia dataset, the Kessler \textit{et al.} \cite{kessler09} (sample combination ``e'') compilation, and analyze it with two light-curve fitters (MLCS2k2  and  SALT2). In \cite{ishida08}, the BAO/CMB constrains upon the model parameters were obtained using only two BAO measurements, whereas here we have used six.  There is no difference between what has been referred to in \cite{ishida08} by the $S_k/D_V$ test, and what is denominated here the BAO/CMB $d_A/D_V$ test.  Our figure \ref{fig4} (left panel) is essentially the same as figure 1 (right panel) of \cite{ishida08}. It is worth mentioning that  $d_A(z_*)/D_V(z_{BAO})$ is equal to $z_{BAO}$ times the ratio $\mathcal{R}(z_*)/\mathcal{A}(z_{BAO})$, where $\mathcal{R}$ is the CMB shift parameter and $\mathcal{A}$ is the BAO scale \cite{eisenstein05}. Notice that this ratio eliminates the dependence on $\Omega_{m0}$ present in the definition of these quantities and, therefore, is less model dependent.

In \cite{ishida08} a tension was found between the results obtained using the \textit{Gold182} dataset and those obtained with the SNLS dataset.  The comparison was performed only for the special case $q_f=-1$ and the origin of the tension was attributed to the datasets themselves. Here, with a more general approach (we do not fix $q_f$), we showed that the origin of the tension might not be the dataset (we used only one) but the light-curve fitter adopted. For flat space, when combined with BAO/CMB, SALT2 tends to favor standard $\Lambda$CDM while MLCS2k2 favors the DGP model (see figure \ref{fig7}). This confirms, with a quite model independent approach, a similar result obtained in \cite{sollerman09} by using an information-criteria model selection. We remark that the discrepancy between the MLCS2k2 and SALT2 light-curve fitters has originally been pointed out in \cite{kessler09}.  Our results are also consistent with those of \cite{kessler09} that considered the more conventional $\Lambda$CDM and flat $w$CDM models. For instance, when combining MLCS2k2 with BAO and CMB measurements they obtained $w= -0.76$ as best-fit (SN Ia sample combination ``e'') for flat $w$CDM, while for SALT2 they got $w=-0.96$, that is in good agreement with $\Lambda$CDM. Interestingly they found that this difference of $\sim 0.2$ in the dark energy equation of state parameter is reduced to $\lesssim 0.04$ if only the SDSS or Nearby+SDSS samples are used. Including ESSENCE gives also consistent results. By considering the mean difference in distance modulus, $\Delta \mu = \mu_{SALT2}-\mu_{MLCS2k2}$, as a function of redshift, for different SNIa samples, they showed that, unlike the other three samples, the mean $\Delta \mu$ is considerably different from zero (and positive) for SNLS and HST samples. Therefore, including these two samples push $w$ to smaller values, as compared to MLCS2k2, when adopting SALT2. Similarly, the same happens in our case. For instance, assuming $q_i=1/2$, $q_f=-1$ and a fixed arbitrary value $z_t$, it can be shown that distances are smaller for DGP ($\tau =1/2$) than for $\Lambda$CDM ($\tau=1/3$) and this qualitatively explains our results. We finally remark that a comprehensive discussion about possible sources for the discrepancy between MLCS2k2 and SALT2 is performed in \cite{kessler09} that trace it to rest-frame ultraviolet modeling and also to the intrinsic color and dust extinction treatment in both fitters.

The main results of this work are displayed in figures \ref{fig6}, \ref{fig7} and table 2. It is now clear that progress in determining the transition redshift and/or the present value of the deceleration parameter, depends crucially on solving the issue of the difference obtained when using these two light-curve fitters. This points out to unresolved systematics as emphasized in \cite{kessler09,sollerman09}. Indeed, in comparison, answering other questions like which parametrization is more adequate to determine $q_0$ or $z_t$, is currently less important. 

A relevant issue, not considered in this work, is to quantify how universal is the assumption $q_i=1/2$. We are currently investigating how good this hypothesis is by taking into account more observables. Another interesting point that deserves further study is how well future SN Ia, BAO and CMB surveys will able to constrain the model parameters. The results of these investigations will be published elsewhere. 

\section*{\textbf{Acknowledgements}}

M.V.S. thanks the Brazilian research agency FAPERJ for support.  I.W. is partially supported by the Brazilian research agency CNPq.

\appendix
\section{Analytic Marginalization over $\mu_0$ in MLCS2k2 and over $\mathcal{M}$ in SALT2 SN data analysis }

\subsection{Considering MLCS2k2 output}
\subsubsection{\bf{Uncorrelated observations}}
Writing explicitly the dependence of $\chi^2_{MLCS2k2}$ on $\mu_0$ in eq. (\ref{eq:chi2_mlcs2k2}), we have
\begin{equation}
\chi^2_{\mathrm{MLCS2k2}}(\mu_0,\boldsymbol{\theta},\sigma_{int}) = \sum_i^N \frac{\{\mu_{i}-5\log[\mathcal{D}_L(z_i;\boldsymbol{\theta})]-\mu_0\}^2}{\sigma_{T,i}^2(\sigma_{int})},
\label{chi2mu}
\end{equation}
Introducing the following auxiliary quantities
\begin{eqnarray}
A(\boldsymbol{\theta},\sigma_{int}) & := & \sum_i^N \frac{\{\mu_{i}-5\log[\mathcal{D}_L(z_i;\boldsymbol{\theta})]\}^2}{\sigma_{T,i}^2(\sigma_{int})}, \\
B(\boldsymbol{\theta},\sigma_{int}) & := & \sum_i^N \frac{\{\mu_{i}-5\log[\mathcal{D}_L(z_i;\boldsymbol{\theta})]\}}{\sigma_{T,i}^2(\sigma_{int})}, \\
C(\sigma_{int}) & := & \sum_i^N \frac{1}{\sigma_{T,i}^2(\sigma_{int})}, \\
D(\sigma_{int}) & := & \prod_i^N \frac{1}{\sigma_{T,i}(\sigma_{int})\sqrt{2\pi}},
\end{eqnarray}
we can rewrite eq. (\ref{chi2mu}) as
\begin{equation}
\chi^2_{\mathrm{MLCS2k2}}(\mu_0,\boldsymbol{\theta},\sigma_{int})=
A(\boldsymbol{\theta},\sigma_{int})-2B(\boldsymbol{\theta},\sigma_{int})\mu_0
+C(\sigma_{int})\mu_0^2,
\end{equation}
consequently, the likelihood, eq. (\ref{likeMLCS2k2}), will be expressed by
\begin{equation}
L(\mu_0,\boldsymbol{\theta},\sigma_{int})=
D(\sigma_{int})\exp\left[-\frac{A(\boldsymbol{\theta},\sigma_{int})}{2}+B(\boldsymbol{\theta},\sigma_{int})\mu_0
-\frac{C(\sigma_{int})}{2}\mu_0^2\right].
\label{likemu1}
\end{equation}
The marginalized likelihood will be
\begin{equation}
L^\ast(\boldsymbol{\theta},\sigma_{int}):=\int_{-\infty}^{+\infty}
L(\mu_0,\boldsymbol{\theta},\sigma_{int})d\mu_0.
\label{marglike}
\end{equation}
Substituting eq. (\ref{likemu1}) in (\ref{marglike}) we obtain
\begin{equation}
L^\ast(\boldsymbol{\theta},\sigma_{int})=
D(\sigma_{int})\sqrt{\frac{2\pi}{C(\sigma_{int})}}\exp\left[-\frac{A(\boldsymbol{\theta},\sigma_{int})}{2}+\frac{B^2(\boldsymbol{\theta},\sigma_{int})}{2C(\sigma_{int})}
\right].
\end{equation}
Finally, we can write
\begin{equation}
{\chi^\ast}^2_{MLCS2k2}(\boldsymbol{\theta},\sigma_{int}):=
-2\ln L^\ast(\boldsymbol{\theta},\sigma_{int})=
A(\boldsymbol{\theta},\sigma_{int})-
\frac{B^2(\boldsymbol{\theta},\sigma_{int})}{C(\sigma_{int})}
-2\ln\left[D(\sigma_{int})\sqrt{\frac{2\pi}{C(\sigma_{int})}}\right]
\end{equation}

\subsubsection{\bf{Correlated observations}}
In the case of correlated observations we have
\begin{equation}
L(\mu_0,\boldsymbol{\theta},\sigma_{int})=
\frac{1}{\sqrt{(2\pi)^N \hbox{det}\boldsymbol{C}}}
\exp\left(-\frac{1}{2}\boldsymbol{X}^T\boldsymbol{C}^{-1}
\boldsymbol{X}\right),
\end{equation}
where, for MLCS2k2 data,
\begin{equation}
X_i=[\mu_i-\mu_{th}(z_i;\mu_0,\boldsymbol{\theta})]=
[\mu_i-5\log\mathcal{D}_L(z_i;\boldsymbol{\theta})-\mu_0].
\end{equation}
Following  \cite{conley2011} we will write
the off-diagonal terms of $\boldsymbol{C}$ as
\begin{equation}
C^{i\neq j}_{ij}=\sum_{k=1}^M\frac{\partial\mu_{th}(z_i;\mu_0,\boldsymbol{\theta})}
{\partial S_k}\frac{\partial\mu_{th}(z_j;\mu_0,\boldsymbol{\theta})}
{\partial S_k}(\Delta S_k)^2,
\end{equation}
where the sum is over $M$ systematics $S_k$ and $\Delta S_k$ is
the uncertainty associated to $S_k$. Given this expression, we find
that $C^{i\neq j}_{ij}$ do not depend on $\mu_0$.

Introducing a new vector
\begin{equation}
\bar{X}_i=[\mu_i-5\log\mathcal{D}_L(z_i;\boldsymbol{\theta})],
\end{equation}
such that
\begin{equation}
\boldsymbol{X}=\bar{\boldsymbol{X}}-\mu_0\vec{\boldsymbol{1}},
\end{equation}
we can write
\begin{equation}
\chi^2=\boldsymbol{X}^T\boldsymbol{C}^{-1}
\boldsymbol{X}=(\bar{\boldsymbol{X}}-\mu_0\vec{\boldsymbol{1}})^T
\boldsymbol{C}^{-1}
(\bar{\boldsymbol{X}}-\mu_0\vec{\boldsymbol{1}}).
\end{equation}
Introducing new auxiliary quantities
\begin{eqnarray}
a(\boldsymbol{\theta},\sigma_{int}) & := & \bar{\boldsymbol{X}}^T\boldsymbol{C}^{-1}
\bar{\boldsymbol{X}}, \\
b(\boldsymbol{\theta},\sigma_{int}) & := & \bar{\boldsymbol{X}}^T\boldsymbol{C}^{-1}
\vec{\boldsymbol{1}}, \\
e(\sigma_{int}) & := & \vec{\boldsymbol{1}}^T\boldsymbol{C}^{-1}\vec{\boldsymbol{1}}, \\
d(\sigma_{int}) & := & \frac{1}{\sqrt{(2\pi)^N \hbox{det}\boldsymbol{C}}},
\end{eqnarray}
we have
\begin{equation}
\chi^2_{\mathrm{MLCS2k2}}(\mu_0,\boldsymbol{\theta},\sigma_{int})=
a(\boldsymbol{\theta},\sigma_{int})-2b(\boldsymbol{\theta},\sigma_{int})\mu_0
+e(\sigma_{int})\mu_0^2,
\end{equation}
and
\begin{equation}
L(\mu_0,\boldsymbol{\theta},\sigma_{int})=
d(\sigma_{int})\exp\left[-\frac{a(\boldsymbol{\theta},\sigma_{int})}{2}+b(\boldsymbol{\theta},\sigma_{int})\mu_0
-\frac{e(\sigma_{int})}{2}\mu_0^2\right].
\label{likemu}
\end{equation}
Finally, marginalizing over $\mu_0$, we have
\begin{equation}
L^\ast(\boldsymbol{\theta},\sigma_{int})=
d(\sigma_{int})\sqrt{\frac{2\pi}{e(\sigma_{int})}}\exp\left[-\frac{a(\boldsymbol{\theta},\sigma_{int})}{2}+\frac{b^2(\boldsymbol{\theta},\sigma_{int})}{2e(\sigma_{int})}
\right].
\end{equation}
and
\begin{equation}
{\chi^\ast}^2_{MLCS2k2}(\boldsymbol{\theta},\sigma_{int})=
a(\boldsymbol{\theta},\sigma_{int})-
\frac{b^2(\boldsymbol{\theta},\sigma_{int})}{e(\sigma_{int})}
-2\ln\left[d(\sigma_{int})\sqrt{\frac{2\pi}{e(\sigma_{int})}}\right]
\end{equation}

\subsection{Considering SALT2 output}
\subsubsection{\bf{Uncorrelated observations}}
Let us write eq. (\ref{eq:chi2SALT2b}) as
\begin{equation}
\chi^2_{\mathrm{SALT2}}(\boldsymbol{\theta},\alpha,\beta,
\mathcal{M},\sigma_{int}) := \sum_i^N \frac{[m_{B,i}^{\ast}-{m_B^*}_{th}(z_i;\boldsymbol{\theta},\alpha,\beta,
\mathcal{M})]^2}{\sigma_{T,i}^2(\alpha,\beta,\sigma_{int})}, \label{chi2_salt2}
\end{equation}
where
\begin{equation}
\sigma_{T,i}^2(\alpha,\beta,\sigma_{int}):=\sigma_{m_{B,i}^{\ast}}^2
+(\sigma^\mu_{z_j})^2+\alpha^2\sigma_{x_{1},i}^2+\beta^2\sigma_{c,i}^2+\sigma_{int}^2.
\end{equation}
Introducing the following auxiliary quantities
\begin{eqnarray}
A'(\boldsymbol{\theta},\alpha,\beta,\sigma_{int}) & := & \sum_i^N \frac{\{m_{B,i}^{\ast}-5\log[\mathcal{D}_L(z_i;\boldsymbol{\theta})]
+\alpha x_{1,i}-\beta c_{i}\}^2}{\sigma_{T,i}^2(\sigma_{int})}, \\
B'(\boldsymbol{\theta},\alpha,\beta,\sigma_{int}) & := & \sum_i^N \frac{\{m_{B,i}^{\ast}-5\log[\mathcal{D}_L(z_i;\boldsymbol{\theta})]
+\alpha x_{1,i}-\beta c_{i}\}}{\sigma_{T,i}^2(\sigma_{int})}, \\
C'(\alpha,\beta,\sigma_{int}) & := & \sum_i^N \frac{1}
{\sigma_{T,i}^2(\alpha,\beta,\sigma_{int})}, \\
D'(\alpha,\beta,\sigma_{int}) & := & \prod_i^N \frac{1}{\sigma_{T,i}(\alpha,\beta,\sigma_{int})\sqrt{2\pi}},
\end{eqnarray}
we can rewrite eq. (\ref{chi2_salt2}) as
\begin{equation}
\chi^2_{\mathrm{SALT2}}(\boldsymbol{\theta},\alpha,\beta,\mathcal{M},\sigma_{int})=
A'(\boldsymbol{\theta},\alpha,\beta,\sigma_{int})-2B'(\boldsymbol{\theta},\alpha,\beta,\sigma_{int})\mathcal{M}
+C'(\alpha,\beta\sigma_{int})\mathcal{M}^2,
\end{equation}
consequently, the likelihood will be expressed by
\begin{equation}
L(\boldsymbol{\theta},\alpha,\beta,\mathcal{M},\sigma_{int})=
D'(\alpha,\beta,\sigma_{int})\exp\left[-\frac{A'(\boldsymbol{\theta}
,\alpha,\beta,\sigma_{int})}{2}+B'(\boldsymbol{\theta},
\alpha,\beta,\sigma_{int})\mathcal{M}
-\frac{C'(\alpha,\beta,\sigma_{int})}{2}\mathcal{M}^2\right].
\label{likemathcal}
\end{equation}
The marginalized likelihood will be
\begin{equation}
L^\ast(\boldsymbol{\theta},\alpha,\beta,\sigma_{int}):=\int_{-\infty}^{+\infty}
L(\boldsymbol{\theta},\alpha,\beta,\mathcal{M},\sigma_{int})d\mathcal{M}.
\label{marglikesalt}
\end{equation}
Using eq. (\ref{likemathcal}) in (\ref{marglikesalt}) we have
\begin{equation}
L^\ast(\boldsymbol{\theta},\alpha,\beta,\sigma_{int})=
D'(\alpha,\beta,\sigma_{int})\sqrt{\frac{2\pi}{C'(\alpha,\beta,\sigma_{int})}}\exp\left[-\frac{A'(\boldsymbol{\theta},\alpha,\beta,\sigma_{int})}{2}+\frac{B'^2(\boldsymbol{\theta},\alpha,\beta,\sigma_{int})}{2C'(\alpha,\beta,\sigma_{int})}
\right].
\end{equation}
Finally, we can write
\begin{eqnarray}
{\chi^\ast}^2_{SALT2}(\boldsymbol{\theta},\alpha,\beta,\sigma_{int}):=
-2\ln L^\ast(\boldsymbol{\theta},\alpha,\beta,\sigma_{int})=\\ \nonumber
=A'(\boldsymbol{\theta},\alpha,\beta,\sigma_{int})-
\frac{B'^2(\boldsymbol{\theta},\alpha,\beta,\sigma_{int})}{C'(\alpha,\beta,\sigma_{int})}
-2\ln\left[D'(\alpha,\beta,\sigma_{int})\sqrt{\frac{2\pi}{C'(\alpha,\beta,\sigma_{int})}}\right]
\end{eqnarray}

\subsubsection{\bf{Correlated observations}}
In the case of correlated observations we just need to choose
\begin{equation}
X_i=[m_{B,i}^{\ast}-m_{th}(\boldsymbol{\theta},\alpha,\beta,\mathcal{M},z_i)]=
[m_{B,i}^{\ast}-5\log\mathcal{D}_L(\boldsymbol{\theta},z_i)+\alpha x_{1,i}
-\beta c_{i}-\mathcal{M}].
\end{equation}
Following \cite{conley2011} again, we will write
the off-diagonal terms of $\boldsymbol{C}$ as
\begin{equation}
C^{i\neq j}_{ij}=\sum_{k=1}^M\frac{\partial m_{th}(\boldsymbol{\theta},\alpha,\beta,\mathcal{M},z_i)}
{\partial S_k}\frac{\partial m_{th}(\boldsymbol{\theta},\alpha,\beta,\mathcal{M},z_j)}
{\partial S_k}(\Delta S_k)^2.
\end{equation}
Now we find
that $C^{i\neq j}_{ij}$ do not depend on $\mathcal{M}$.

Introducing a new vector
\begin{equation}
\bar{X}_i=[m_{B,i}^{\ast}-5\log\mathcal{D}_L(\boldsymbol{\theta},z_i)+\alpha x_{1,i}
-\beta c_{i}],
\end{equation}
such that
\begin{equation}
\boldsymbol{X}=\bar{\boldsymbol{X}}-\mathcal{M}\vec{\boldsymbol{1}},
\end{equation}
and introducing the auxiliary quantities
\begin{eqnarray}
a'(\boldsymbol{\theta},\alpha,\beta,\sigma_{int}) & := & \bar{\boldsymbol{X}}^T\boldsymbol{C}^{-1}
\bar{\boldsymbol{X}}, \\
b'(\boldsymbol{\theta},\alpha,\beta,\sigma_{int}) & := & \bar{\boldsymbol{X}}^T\boldsymbol{C}^{-1}
\vec{\boldsymbol{1}}, \\
e'(\alpha,\beta,\sigma_{int}) & := & \vec{\boldsymbol{1}}^T\boldsymbol{C}^{-1}\vec{\boldsymbol{1}}, \\
d'(\alpha,\beta,\sigma_{int}) & := & \frac{1}{\sqrt{(2\pi)^N \hbox{det}\boldsymbol{C}}},
\end{eqnarray}
we have
\begin{equation}
\chi^2_{\mathrm{SALT2}}(\boldsymbol{\theta},\alpha,\beta,\mathcal{M},\sigma_{int})=
a'(\boldsymbol{\theta},\alpha,\beta,\sigma_{int})-2b'(\boldsymbol{\theta},\alpha,\beta,\sigma_{int})\mathcal{M}
+e'(\alpha,\beta,\sigma_{int})\mathcal{M}^2,
\end{equation}
and
\begin{equation}
L(\boldsymbol{\theta},\alpha,\beta,\mathcal{M},\sigma_{int})=
d'(\alpha,\beta,\sigma_{int})\exp\left[-\frac{a'(\boldsymbol{\theta},\alpha,\beta,\sigma_{int})}{2}+b'(\boldsymbol{\theta},\alpha,\beta,\sigma_{int})\mathcal{M}
-\frac{e'(\sigma_{int})}{2}\mathcal{M}^2\right].
\label{likemsalt2}
\end{equation}
Finally, marginalizing over $\mathcal{M}$, we have
\begin{equation}
L^\ast(\boldsymbol{\theta},\alpha,\beta,\sigma_{int})=
d'(\alpha,\beta,\sigma_{int})\sqrt{\frac{2\pi}{e'(\alpha,\beta,\sigma_{int})}}\exp\left[-\frac{a'(\boldsymbol{\theta},\alpha,\beta,\sigma_{int})}{2}+\frac{b'^2(\boldsymbol{\theta},\alpha,\beta,\sigma_{int})}{2e'(\alpha,\beta,\sigma_{int})}
\right].
\end{equation}
and
\begin{equation}
{\chi^\ast}^2_{SALT2}(\boldsymbol{\theta},\alpha,\beta,\sigma_{int})=
a'(\boldsymbol{\theta},\alpha,\beta,\sigma_{int})-
\frac{b'^2(\boldsymbol{\theta},\alpha,\beta,\sigma_{int})}{e'(\alpha,\beta,\sigma_{int})}
-2\ln\left[d'(\alpha,\beta,\sigma_{int})\sqrt{\frac{2\pi}{e'(\alpha,\beta,\sigma_{int})}}\right]
\end{equation}

\bibliography{arXiv_qpar}

\begin{thebibliography}{20}
\expandafter\ifx\csname natexlab\endcsname\relax\def\natexlab#1{#1}\fi

\bibitem{Riess1998} 
Riess A G \textit{et al.}, \emph{Observational evidence
from supernovae for an accelerating universe and a cosmological constant}, 1998 \emph{Astron. J.} {\bf 116}, 1009 [astro-ph/9805201]; \\ 
Perlmutter S \textit{et al.}, \emph{Measurements of
$\Omega$ and $\Lambda$ from 42 high-redshift supernovae}, 1999 \emph{Astrophys. J.}
\textbf{517}, 565 [astro-ph/9812133].

\bibitem{Howell2010} Howell D A, \emph{A review of type Ia supernovae as stellar
endpoints and cosmological tools}, 2011 \emph{Nature Communications} \textbf{2}, 350 [arXiv:1011.0441].

\bibitem{Wood-Vasey2007} Wood-Vasey W M \textit{et al.}, \emph{Observational constraints on the nature of dark energy: first cosmological results from the ESSENCE supernova survey}, 2007 \emph{Astrophys. J.} \textbf{666}, 694 [astro-ph/0701041]; \\ Kowalski M \textit{et al.}, \emph{Improved cosmological constraints from new, old and combined supernova datasets}, 2008  \emph{Astrophys. J.} \textbf{686}, 749 [arXiv:0804.4142];\\ Hicken M \textit{et al.}, \emph{Improved dark energy constraints from $\sim$ 100 new CfA supernova type Ia light curves}, 2009 
\emph{Astrophys. J.} \textbf{700}, 1097 [arXiv:0901.4804].

\bibitem{conley2011} Conley A \textit{et al.}, \emph{Supernova constraints and systematic uncertainties from the first 3 years of the Supernova Legacy Survey}, 2011 \emph{Astrophys. J. Suppl.} \textbf{192}, 1 [arXiv:1104.1443].

\bibitem{kessler09} Kessler R \textit{et al.}, \emph{First-year Sloan Digital Skt Survey-II supernova results: Hubble diagram and cosmological parameters}, 2009 
\emph{Astrophys. J. Suppl.} \textbf{185}, 32 [arXiv:0908.4274].

\bibitem{Amanullah2010} Amanullah R \textit{et al.}, \emph{Spectra and Hubble Space Telescope light curves of six type Ia supernovae at $0.511< z< 1.12$ and the Union2 compilation}, 2010 \emph{Astrophys. J.} \textbf{716}, 712 [arXiv:1004.1711].

\bibitem{blake11} Blake C \textit{et al.}, \emph{The WiggleZ Dark Energy Survey: mapping the distance-redshift relation with baryon acoustic oscillations}, 2011 \emph{Mon. Not. R. Astron. Soc.} \textbf{418}, 1707 [arXiv:1108.2635].

\bibitem{percival10} Percival W J \textit{et al.}, \emph{Baryon Acoustic Oscillations in the Sloan Digital Sky Survey Data Release 7 Galaxy Sample}, 2010 \emph{Mon. Not. R. Astron. Soc.} \textbf{401}, 2148 [arXiv:0907.1660].

\bibitem{kinematic} Turner M S  and Riess  A G , \emph{Do SNe Ia provide direct evidence for past deceleration of the universe?}, 2002 \emph{Astrophys. J.} \textbf{569}, 18 [astro-ph/0106051]; 
\\ Riess A G  \textit{et al.}, \emph{Type Ia supernova discoveries at $z > 1$ from the Hubble Space Telescope: Evidence for past deceleration and constraints on dark energy evolution}, 2004 \emph{Astrophys. J.} \textbf{607},
665 [astro-ph/0402512];\\ Elgaroy O and Multamaki T, \emph{Bayesian analysis of friedmannless cosmologies} 2006 \emph{J. Cosmol. Astropart. Phys.}
\textbf{JCAP09(2006)002} [astro-ph/0603053]; 
 \\ Gong Y and Wang A, \emph{Reconstruction of the deceleration parameter and the equation of state of dark energy}, 2007 \emph{Phys. Rev. D};
\textbf{75}, 043520 [astro-ph/0612196];\\ Rapetti D, Allen S W, Amin M A and
Blandford R G, \emph{A kinematical approach to dark energy studies}, 2007 \emph{Mon. Not. R. Astron. Soc.} \textbf{375}, 1510 [astro-ph/0605683];\\ Cunha J V and Lima J A S, \emph{Transition Redshift: New Kinematic Constraints from Supernovae}, 2008 
\emph{Mon. Not. R. Astron. Soc.} \textbf{390}, 210 [arXiv:0805.1261];\\ Mortsell E and Clarkson C, \emph{Model independent constraints on the cosmological expansion rate}, 2009 \emph{J. Cosmol. Astropart. Phys.} \textbf{JCAP01(2009)044} [arXiv:0811.0981];\\ Xu L, Li W and Lu J, \emph{Constraints on Kinematic Model from Recent Cosmic Observations: SN Ia, BAO and Observational Hubble Data}, 2009 \emph{J. Cosmol. Astropart. Phys.} \textbf{JCAP07(2009)031} [arXiv:0905.4552];\\
Santos B, Carvalho J C and Alcaniz J S, \emph{Current constraints on the epoch of cosmic acceleration}, 2011 \emph{Astroparticle Physics} \textbf{35}, 17 [arXiv:1009.2733]; \\
Nair R, Jhingan S and Jain D, \emph{Cosmokinetics: A joint analysis of Standard Candles, Rulers and Cosmic Clocks} 2012 \emph{J. Cosmol. Astropart. Phys.}
\textbf{JCAP01(2012)018} [arXiv:1109.4574].

\bibitem{ishida08} Ishida E E O, Reis R R R, Toribio A V, and Waga I, \emph{When did cosmic acceleration start? How fast was the transition?}, 2008 \emph{Astroparticle Physics} \textbf{28}, 547 [arXiv:0706.0546].

\bibitem{chiba98} Chiba T and Nakamura T, \emph{The Luminosity distance, the equation of state, and the geometry of the universe}, 1998 \emph{Prog. Theor. Phys.} \textbf{100}, 1077 [astro-ph/9808022];\\ Sahni V, \emph{The Cosmological constant problem and quintessence}, 2002 \emph{Class. Quant. Grav.} \textbf{19}, 3435 [astro-ph/0202076].

\bibitem{dvali00} Dvali T M, Gabadadze G, Porrati M, \emph{4-D gravity on a brane in 5-D Minkowski space}, 2000 \emph{Phys. Lett. B} \textbf{485}, 208 [hep-th/0005016].

\bibitem{kamenshchik01} Kamenshchik A Y, Moschella U, Pasquier V, \emph{An Alternative to quintessence}, 2001 \emph{Phys. Lett. B}
\textbf{511}, 8 [gr-qc/0103004].

\bibitem{bilic01}
Bilic N, Tupper G B and Viollier R D, \emph{Unification of dark matter and dark energy: The Inhomogeneous Chaplygin gas}, 2002 \emph{Phys. Lett. B} \textbf{535}, 17 [astro-ph/0111325].

\bibitem{bento02}
Bento M, Bertolami O and Sen A, \emph{Generalized Chaplygin gas, accelerated expansion and dark energy matter unification}, 2002 \emph{Phys. Rev. D} 66, 043507 [gr-qc/0202064].

\bibitem{makler02} Makler M, de Oliveira S Q and Waga I, \emph{Constraints on the generalized Chaplygin gas from supernovae observations}, 2003 \emph{Phys. Lett. B} \textbf{555}, 1 [astro-ph/0209486].

\bibitem{gondolo03} Gondolo P and Freese K, \emph{Fluid interpretation of cardassian expansion}, 2003 \emph{Phys. Rev. D} \textbf{68}, 063509 [hep-ph/0209322].

\bibitem{wang03} Wang Y, Freese K, Gondolo P and Lewis M, \emph{Future Type IA supernova data as tests of dark energy from modified Friedmann equations}, 2003 \emph{ Astrophys. J.} \textbf{594}, 25 [astro-ph/0302064].


\bibitem{beutler11} Beutler F \textit{et al.}, \emph{The 6dF Galaxy Survey: Baryon Acoustic Oscillations and the Local Hubble Constant}, 2011 \emph{Mon. Not. R. Astron. Soc.} \textbf{416}, 3017 [arXiv:1106.3366].

\bibitem{komatsu11} Komatsu E \textit{et al.}, \emph{Seven-Year Wilkinson Microwave Anisotropy Probe (WMAP) Observations: Cosmological Interpretation}, 2011 \emph{Astrophys. J. Suppl.} \textbf{192}, 18 [arXiv:1001.4538].

\bibitem{jarosik11} Jarosik N \textit{et al.}, \emph{Seven-Year Wilkinson Microwave Anisotropy Probe (WMAP) Observations: Sky Maps, Systematic Errors, and Basic Results}, 2011 \emph{Astrophys. J. Suppl.} \textbf{192}, 14 [arXiv:1001.4744].


\bibitem{riess07} Riess A G \textit{et al.}, \emph{New Hubble Space Telescope Discoveries of Type Ia Supernovae at $z\geq1$: Narrowing Constraints on the Early Behavior of Dark Energy}, 2007 \emph{Astrophys. J.} \textbf{659}, 98 [astro-ph/0611572].

\bibitem{astier05}
Astier P \textit{et al.}, \emph{The Supernova Legacy Survey: measurement of $\Omega_M$, $\Omega_{\Lambda}$ and $w$
from the first year data set}, 2006 \emph{Astron. and Astrophys.}, \textbf{447}, 31 [astro-ph/0510447].


\bibitem{guy05}
Guy J, Astier P, Nobili S, Regnault N and Pain R, \emph{SALT: a spectral adaptive light curve template
for type Ia supernovae}, 2005 \emph{Astron. and
Astrophys.} \textbf{443}, 781 [astro-ph/0506583].


\bibitem{jha07}
Jha S, Riess A G and Kirshner R P , \emph{Improved distances to type Ia supernovae with multicolor light-curve shapes: MLCS2k2}, 2007 \emph{Astrophys. J.}, \textbf{659},
122 [astro-ph/0612666].


\bibitem{guy07} Guy J \textit{et al.}, \emph{SALT2: using distant supernovae to improve the use
of type Ia supernovae as distance indicators}, 2007 \emph{Astron. and Astrophys.} \textbf{466}, 11 [astro-ph/0701828].

\bibitem{lago11} Lago B L \textit{et al.}, \emph{Type Ia supernova parameter estimation: a comparison of two approaches using current datasets}, 2011 \emph{Preprint} arXiv:1104.2874.

\bibitem{kim11} Kim A G, \emph{Type Ia Supernova Intrinsic Magnitude Dispersion and the Fitting of Cosmological Parameters}, 2011 \emph{Publications of the Astronomical Society of the Pacific} \textbf{123}, 230 [arXiv:1101.3513].


\bibitem{sollerman09} Sollerman J \textit{et al.}, \emph{First-year Sloan Digital Sky Survey-II (SDSS-II) supernova results: constraints on non-standard cosmological models}, 2009 \emph{Astrophys. J.} \textbf{703}, 1374 [arXiv:0908.4276].

\bibitem{hu96} Hu W and Sugiyama N, \emph{Small scale cosmological perturbations: An Analytic approach}, 1996 \emph{Astrophys. J.} \textbf{471}, 542 [astro-ph/9510117].

\bibitem{eisenstein05} Eisenstein D, \emph{Detection of the baryon acoustic peak in the large-scale correlation function of SDSS luminous red galaxies}, 2005 \emph{Astrophys. J.} \textbf{633}, 560 [astro-ph/0501171].

\bibitem{mcmc} Gregory P C, \emph{Bayesian Logical Data Analysis for the Physical Sciences: A Comparative Approach with Mathematica Support}, 2005, 
Cambridge University Press; \\ Gamerman D and Lopes H F, \emph{Markov Chain Monte Carlo: Stochastic Simulation for Bayesian Inference}, 2006, Chapman \& Hall/CRC Texts in Statistical Science.

\end{thebibliography}

\end{document}